\documentclass[12pt]{article}
\usepackage{lineno}
\usepackage{epsfig}
\usepackage{epstopdf}
\usepackage{amsmath}
\usepackage{hhline}
\usepackage{amssymb}
\usepackage{mathptmx}
\usepackage[scaled=.92]{helvet}
\usepackage{courier}
\usepackage{cite}
\usepackage{rotating}
\usepackage{url}
\usepackage{multirow}
\usepackage{bm} 

\newlength{\dinwidth}
\newlength{\dinmargin}
\def\Journal#1#2#3#4{{#1} {\bf #2}, (#3) #4}

\def\EPJC{{Eur. Phys. J.} {\bf C}}

\def\be{\begin{equation}}
\def\ee{\end{equation}}
\def\bea{\begin{eqnarray}}
\def\eea{\end{eqnarray}}
\def\etal{{\it et~al.}}

\begin{document}  
\newcommand{\pom}{{I\!\!P}}
\newcommand{\slowpi}{\pi_{\mathit{slow}}}
\newcommand{\fiidiii}{F_2^{D(3)}}
\newcommand{\fiidiiiarg}{\fiidiii\,(\beta,\,Q^2,\,x)}
\newcommand{\n}{1.19\pm 0.06 (stat.) \pm0.07 (syst.)}
\newcommand{\nz}{1.30\pm 0.08 (stat.)^{+0.08}_{-0.14} (syst.)}
\newcommand{\fiidiiiful}{F_2^{D(4)}\,(\beta,\,Q^2,\,x,\,t)}
\newcommand{\fiipom}{\tilde F_2^D}
\newcommand{\ALPHA}{1.10\pm0.03 (stat.) \pm0.04 (syst.)}
\newcommand{\ALPHAZ}{1.15\pm0.04 (stat.)^{+0.04}_{-0.07} (syst.)}
\newcommand{\fiipomarg}{\fiipom\,(\beta,\,Q^2)}
\newcommand{\pomflux}{f_{\pom / p}}
\newcommand{\nxpom}{1.19\pm 0.06 (stat.) \pm0.07 (syst.)}
\newcommand {\gapprox}
   {\raisebox{-0.7ex}{$\stackrel {\textstyle>}{\sim}$}}
\newcommand {\lapprox}
   {\raisebox{-0.7ex}{$\stackrel {\textstyle<}{\sim}$}}
\def\gsim{\,\lower.25ex\hbox{$\scriptstyle\sim$}\kern-1.30ex%
\raise 0.55ex\hbox{$\scriptstyle >$}\,}
\def\lsim{\,\lower.25ex\hbox{$\scriptstyle\sim$}\kern-1.30ex%
\raise 0.55ex\hbox{$\scriptstyle <$}\,}
\newcommand{\pomfluxarg}{f_{\pom / p}\,(x_\pom)}
\newcommand{\dsf}{\mbox{$F_2^{D(3)}$}}
\newcommand{\dsfva}{\mbox{$F_2^{D(3)}(\beta,Q^2,x_{I\!\!P})$}}
\newcommand{\dsfvb}{\mbox{$F_2^{D(3)}(\beta,Q^2,x)$}}
\newcommand{\dsfpom}{$F_2^{I\!\!P}$}
\newcommand{\gap}{\stackrel{>}{\sim}}
\newcommand{\lap}{\stackrel{<}{\sim}}
\newcommand{\fem}{$F_2^{em}$}
\newcommand{\tsnmp}{$\tilde{\sigma}_{NC}(e^{\mp})$}
\newcommand{\tsnm}{$\tilde{\sigma}_{NC}(e^-)$}
\newcommand{\tsnp}{$\tilde{\sigma}_{NC}(e^+)$}
\newcommand{\st}{$\star$}
\newcommand{\sst}{$\star \star$}
\newcommand{\ssst}{$\star \star \star$}
\newcommand{\sssst}{$\star \star \star \star$}

\newcommand{\tw}{\theta_W}
\newcommand{\sw}{\sin{\theta_W}}
\newcommand{\cw}{\cos{\theta_W}}
\newcommand{\sww}{\sin^2{\theta_W}}
\newcommand{\cww}{\cos^2{\theta_W}}
\newcommand{\trm}{m_{\perp}}
\newcommand{\trp}{p_{\perp}}
\newcommand{\trmm}{m_{\perp}^2}
\newcommand{\trpp}{p_{\perp}^2}
\newcommand{\alp}{\alpha_s}

\newcommand{\alps}{\alpha_s}
\newcommand{\sqrts}{$\sqrt{s}$}
\newcommand{\LO}{$O(\alpha_s^0)$}
\newcommand{\Oa}{$O(\alpha_s)$}
\newcommand{\Oaa}{$O(\alpha_s^2)$}
\newcommand{\PT}{p_{\perp}}
\newcommand{\JPSI}{J/\psi}
\newcommand{\sh}{\hat{s}}
\newcommand{\uh}{\hat{u}}
\newcommand{\MP}{m_{J/\psi}}
\newcommand{\PO}{I\!\!P}
\newcommand{\xbj}{x}
\newcommand{\xpom}{x_{\PO}}
\newcommand{\ttbs}{\char'134}
\newcommand{\xpomlo}{3\times10^{-4}}  
\newcommand{\xpomup}{0.05}  
\newcommand{\dgr}{^\circ}
\newcommand{\pbarnt}{\,\mbox{{\rm pb$^{-1}$}}}
\newcommand{\gev}{\,\mbox{GeV}}
\newcommand{\WBoson}{\mbox{$W$}}
\newcommand{\fbarn}{\,\mbox{{\rm fb}}}
\newcommand{\fbarnt}{\,\mbox{{\rm fb$^{-1}$}}}
%
%
\newcommand{\qsq}{\ensuremath{Q^2} }
\newcommand{\gevsq}{\ensuremath{\mathrm{GeV}^2} }
\newcommand{\et}{\ensuremath{E_t^*} }
\newcommand{\rap}{\ensuremath{\eta^*} }
\newcommand{\gp}{\ensuremath{\gamma^*}p }
\newcommand{\dsiget}{\ensuremath{{\rm d}\sigma_{ep}/{\rm d}E_t^*} }
\newcommand{\dsigrap}{\ensuremath{{\rm d}\sigma_{ep}/{\rm d}\eta^*} }

\begin{titlepage}

\noindent
\noindent
\begin{flushleft}
{\tt DESY 15-037    \hfill    ISSN 0418-9833} \\
{\tt March 2015}                  \\

\end{flushleft}

\vspace{1cm}

\vspace{3.5cm}
\begin{center}
\begin{Large}
\boldmath
{\bf Combination of Differential $\pmb{D^{*\pm}}$ Cross-Section Measurements in Deep-Inelastic $\bm{e\!\!\;p}$ Scattering at HERA 
}
\unboldmath
\vspace{2cm}

\end{Large}§
\end{center}

\vspace{2cm}
\def\gev{\rm GeV}
\def\ie{\it i.e.}
\def\etal{\hbox{$\it et~al.$}}
\def\clb#1 {(#1 Coll.),}
\hyphenation{do-mi-nant
}

\begin{abstract}
\noindent

H1 and ZEUS have 
published single-differential cross sections for inclusive $D^{*\pm}$-meson production in deep-inelastic $ep$ scattering at HERA from their respective final data sets.
These cross sections are combined in the common visible phase-space region of photon virtuality 
$Q^2 > 5$~GeV$^2$, electron inelasticity $0.02<y<0.7$ and the $D^{*\pm}$ meson's transverse momentum 
$p_T(D^*)>1.5$~GeV and pseudorapidity $|\eta(D^*)|<1.5$.
The combination procedure takes into account all correlations, yielding significantly reduced experimental uncertainties.
Double-differential cross sections 
${\rm d}^2\sigma/{{\rm d}Q^2{\rm d}y}$ are combined with earlier $D^{*\pm}$ data, 
extending the kinematic range down to $Q^2>1.5$ GeV$^2$. 
Perturbative next-to-leading-order QCD predictions are compared to the results.

\end{abstract}
\vspace{1.5cm}
\begin{center}
Submitted to JHEP
\end{center}
\end{titlepage}
\newpage
{\Large The H1 and ZEUS Collaborations}

{\small\raggedright
H.~Abramowicz$^{\mathrm{49},\mathrm{a1}}$,
I.~Abt$^{\mathrm{36}}$,
L.~Adamczyk$^{\mathrm{22}}$,
M.~Adamus$^{\mathrm{58}}$,
V.~Andreev$^{\mathrm{34}}$,
S.~Antonelli$^{\mathrm{6}}$,
V.~Aushev$^{\mathrm{25},\mathrm{26},\mathrm{b21}}$,
Y.~Aushev$^{\mathrm{26},\mathrm{a2},\mathrm{b21}}$,
A.~Baghdasaryan$^{\mathrm{60}}$,
K.~Begzsuren$^{\mathrm{55}}$
% ...

\newpage
\footnotesize\begin{description}\setlength{\parsep}{0em}\setlength{\itemsep}{0em}
\item[$^{1}$] 
 I. Physikalisches Institut der RWTH, Aachen, Germany
\item[$^{2}$] 
 {Institute of Physics and Technology of Ministry of Education and Science of Kazakhstan, Almaty, Kazakhstan}
\item[$^{3}$] 
 Inter-University Institute for High Energies ULB-VUB, Brussels and Universiteit Antwerpen, Antwerpen, Belgium$^{\mathrm{b1}}$
\item[$^{4}$] 
 School of Physics and Astronomy, University of Birmingham, Birmingham, UK$^{\mathrm{b2}}$
\item[$^{5}$] 
 {INFN Bologna, Bologna, Italy{}}$^{\mathrm{b3}}$
\item[$^{6}$] 
 {University and INFN Bologna, Bologna, Italy{}}$^{\mathrm{b3}}$
\item[$^{7}$] 
 {Physikalisches Institut der Universit\"at Bonn, Bonn, Germany{}}$^{\mathrm{b4}}$
\item[$^{8}$] 
 National Institute for Physics and Nuclear Engineering (NIPNE) , Bucharest, Romania$^{\mathrm{b5}}$
\item[$^{9}$] 
 {Panjab University, Department of Physics, Chandigarh, India}
\item[$^{10}$] 
 {Calabria University, Physics Department and INFN, Cosenza, Italy{}}$^{\mathrm{b3}}$
\item[$^{11}$] 
 STFC, Rutherford Appleton Laboratory, Didcot, Oxfordshire, UK$^{\mathrm{b2}}$
\item[$^{12}$] 
 Institut f\"ur Physik, TU Dortmund, Dortmund, Germany$^{\mathrm{b6}}$
\item[$^{13}$] 
 Joint Institute for Nuclear Research, Dubna, Russia
\item[$^{14}$] 
 CEA, DSM/Irfu, CE-Saclay, Gif-sur-Yvette, France
\item[$^{15}$] 
 {School of Physics and Astronomy, University of Glasgow, Glasgow, United Kingdom{}}$^{\mathrm{b2}}$
\item[$^{16}$] 
 Institut f\"ur Experimentalphysik, Universit\"at Hamburg, Hamburg, Germany$^{\mathrm{b6},\mathrm{b7}}$
\item[$^{17}$] 
 {Deutsches Elektronen-Synchrotron DESY, Hamburg, Germany}
\item[$^{18}$] 
 Physikalisches Institut, Universit\"at Heidelberg, Heidelberg, Germany$^{\mathrm{b6}}$
\item[$^{19}$] 
 Kirchhoff-Institut f\"ur Physik, Universit\"at Heidelberg, Heidelberg, Germany$^{\mathrm{b6}}$
\item[$^{20}$] 
 Institute of Experimental Physics, Slovak Academy of Sciences, Ko\v{s}ice, Slovak Republic$^{\mathrm{b8}}$
\item[$^{21}$] 
 {The Henryk Niewodniczanski Institute of Nuclear Physics, Polish Academy of \ Sciences, Krakow, Poland{}}$^{\mathrm{b9},\mathrm{b15}}$
\item[$^{22}$] 
 {AGH-University of Science and Technology, Faculty of Physics and Applied Computer Science, Krakow, Poland{}}$^{\mathrm{b9}}$
\item[$^{23}$] 
 {Department of Physics, Jagellonian University, Krakow, Poland}
\item[$^{24}$] 
 {National Centre for Particle Physics, Universiti Malaya, 50603 Kuala Lumpur, Malaysia{}}$^{\mathrm{b10}}$
\item[$^{25}$] 
 {Institute for Nuclear Research, National Academy of Sciences, Kyiv, Ukraine}
\item[$^{26}$] 
 {Department of Nuclear Physics, National Taras Shevchenko University of Kyiv, Kyiv, Ukraine}
\item[$^{27}$] 
 Department of Physics, University of Lancaster, Lancaster, UK$^{\mathrm{b2}}$
\item[$^{28}$] 
 Department of Physics, University of Liverpool, Liverpool, UK$^{\mathrm{b2}}$
\item[$^{29}$] 
 School of Physics and Astronomy, Queen Mary, University of London, London, UK$^{\mathrm{b2}}$
\item[$^{30}$] 
 {Physics and Astronomy Department, University College London, London, United Kingdom{}}$^{\mathrm{b2}}$
\item[$^{31}$] 
 Aix Marseille Universit\'{e}, CNRS/IN2P3, CPPM UMR 7346, 13288 Marseille, France
\item[$^{32}$] 
 Departamento de Fisica Aplicada, CINVESTAV, M\'erida, Yucat\'an, M\'exico$^{\mathrm{b11}}$
\item[$^{33}$] 
 Institute for Theoretical and Experimental Physics, Moscow, Russia$^{\mathrm{b12}}$
\item[$^{34}$] 
 Lebedev Physical Institute, Moscow, Russia
\item[$^{35}$] 
 {Lomonosov Moscow State University, Skobeltsyn Institute of Nuclear Physics, Moscow, Russia{}}$^{\mathrm{b13}}$
\item[$^{36}$] 
 {Max-Planck-Institut f\"ur Physik, M\"unchen, Germany}
\item[$^{37}$] 
 {Department of Physics, York University, Ontario, Canada M3J 1P3{}}$^{\mathrm{b14}}$
\item[$^{38}$] 
 LAL, Universit\'e Paris-Sud, CNRS/IN2P3, Orsay, France
\item[$^{39}$] 
 {Department of Physics, University of Oxford, Oxford, United Kingdom{}}$^{\mathrm{b2}}$
\item[$^{40}$] 
 {INFN Padova, Padova, Italy{}}$^{\mathrm{b3}}$
\item[$^{41}$] 
 {Dipartimento di Fisica e Astronomia dell' Universit\`a and INFN, Padova, Italy{}}$^{\mathrm{b3}}$
\item[$^{42}$] 
 LLR, Ecole Polytechnique, CNRS/IN2P3, Palaiseau, France
\item[$^{43}$] 
 Faculty of Science, University of Montenegro, Podgorica, Montenegro$^{\mathrm{b16}}$
\item[$^{44}$] 
 Institute of Physics, Academy of Sciences of the Czech Republic, Praha, Czech Republic$^{\mathrm{b17}}$
\item[$^{45}$] 
 Faculty of Mathematics and Physics, Charles University, Praha, Czech Republic$^{\mathrm{b17}}$
\item[$^{46}$] 
 {Department of Physics, McGill University, Montr\'eal, Qu\'ebec, Canada H3A 2T8{}}$^{\mathrm{b14}}$
\item[$^{47}$] 
 {Department of Particle Physics and Astrophysics, Weizmann Institute, Rehovot, Israel}
\item[$^{48}$] 
 Institute for Nuclear Research and Nuclear Energy, Sofia, Bulgaria
\item[$^{49}$] 
 {Raymond and Beverly Sackler Faculty of Exact Sciences, School of Physics, \ Tel Aviv University, Tel Aviv, Israel{}}$^{\mathrm{b18}}$
\item[$^{50}$] 
 {Polytechnic University, Tokyo, Japan{}}$^{\mathrm{b19}}$
\item[$^{51}$] 
 {Department of Physics, Tokyo Institute of Technology, Tokyo, Japan{}}$^{\mathrm{b19}}$
\item[$^{52}$] 
 {Universit\`a di Torino and INFN, Torino, Italy{}}$^{\mathrm{b3}}$
\item[$^{53}$] 
 {Universit\`a del Piemonte Orientale, Novara, and INFN, Torino, Italy{}}$^{\mathrm{b3}}$
\item[$^{54}$] 
 {Institute of Particle and Nuclear Studies, KEK, Tsukuba, Japan{}}$^{\mathrm{b19}}$
\item[$^{55}$] 
 Institute of Physics and Technology of the Mongolian Academy of Sciences, Ulaanbaatar, Mongolia
\item[$^{56}$] 
 Paul Scherrer Institut, Villigen, Switzerland
\item[$^{57}$] 
 {Faculty of Physics, University of Warsaw, Warsaw, Poland}
\item[$^{58}$] 
 {National Centre for Nuclear Research, Warsaw, Poland}
\item[$^{59}$] 
 Fachbereich C, Universit\"at Wuppertal, Wuppertal, Germany
\item[$^{60}$] 
 Yerevan Physics Institute, Yerevan, Armenia
\item[$^{61}$] 
 {Meiji Gakuin University, Faculty of General Education, Yokohama, Japan{}}$^{\mathrm{b19}}$
\item[$^{62}$] 
 {Deutsches Elektronen-Synchrotron DESY, Zeuthen, Germany}
\item[$^{63}$] 
 Institut f\"ur Teilchenphysik, ETH, Z\"urich, Switzerland$^{\mathrm{b20}}$
\item[$^{64}$] 
 Physik-Institut der Universit\"at Z\"urich, Z\"urich, Switzerland$^{\mathrm{b20}}$
\item[$^{\dagger}$] 
 {Deceased}
\end{description}
\medskip\goodbreak
\begin{description}\setlength{\parsep}{0em}\setlength{\itemsep}{0em}\item[$^{\mathrm{a1}}$] 
 Also at Max Planck Institute for Physics, Munich, Germany, External Scientific Member
\item[$^{\mathrm{a2}}$] 
 Member of National Technical University of Ukraine, Kyiv Polytechnic Institute, Kyiv, Ukraine
\item[$^{\mathrm{a3}}$] 
 Also at Moscow Institute of Physics and Technology, Moscow, Russia
\item[$^{\mathrm{a4}}$] 
 Also at \L\'{o}d\'{z} University, Poland
\item[$^{\mathrm{a5}}$] 
 Now at Rockefeller University, New York, NY 10065, USA
\item[$^{\mathrm{a6}}$] 
 Also at Rechenzentrum, Universit\"at Wuppertal, Wuppertal, Germany
\item[$^{\mathrm{a7}}$] 
 Alexander von Humboldt Professor; also at DESY and University of Oxford
\item[$^{\mathrm{a8}}$] 
 Now at School of Physics and Astronomy, University of Birmingham, UK
\item[$^{\mathrm{a9}}$] 
 Also at IPNL, Universit\'e Claude Bernard Lyon 1, CNRS/IN2P3, Villeurbanne, France
\item[$^{\mathrm{a10}}$] 
 Now at European X-ray Free-Electron Laser facility GmbH, Hamburg, Germany
\item[$^{\mathrm{a11}}$] 
 Now at Fermilab, Chicago, United States
\item[$^{\mathrm{a12}}$] 
 Now at University of Liverpool, United Kingdom
\item[$^{\mathrm{a13}}$] 
 Now at DESY ATLAS group
\item[$^{\mathrm{a14}}$] 
 Now at LNF, Frascati, Italy
\item[$^{\mathrm{a15}}$] 
 Also at Agensi Nuklear Malaysia, 43000 Kajang, Bangi, Malaysia
\item[$^{\mathrm{a16}}$] 
 Now at CERN, Geneva, Switzerland
\item[$^{\mathrm{a17}}$] 
 Member of \L\'{o}d\'{z} University, Poland
\item[$^{\mathrm{a18}}$] 
 Also at Department of Physics, University of Toronto, Toronto, Ontario, Canada M5S 1A7
\item[$^{\mathrm{a19}}$] 
 Also at LAPP, Universit\'e de Savoie, CNRS/IN2P3, Annecy-le-Vieux, France
\item[$^{\mathrm{a20}}$] 
 Now at BNL, USA
\item[$^{\mathrm{a21}}$] 
 Member of National University of Kyiv - Mohyla Academy, Kyiv, Ukraine
\item[$^{\mathrm{a22}}$] 
 Also at University of Tokyo, Japan
\item[$^{\mathrm{a23}}$] 
 Now at Polish Air Force Academy in Deblin
\item[$^{\mathrm{a24}}$] 
 Also at Ulaanbaatar University, Ulaanbaatar, Mongolia
\item[$^{\mathrm{a25}}$] 
 Also at Universit\"{a}t Hamburg and supported by DESY and the Alexander von Humboldt Foundation
\item[$^{\mathrm{a26}}$] 
 Now at Kobe University, Japan
\end{description}
\medskip\goodbreak
\begin{description}\setlength{\parsep}{0em}\setlength{\itemsep}{0em}\item[$^{\mathrm{b1}}$] 
 Supported by FNRS-FWO-Vlaanderen, IISN-IIKW and IWT and by Interuniversity Attraction Poles Programme, Belgian Science Policy
\item[$^{\mathrm{b2}}$] 
 Supported by the UK Science and Technology Facilities Council, and formerly by the UK Particle Physics and Astronomy Research Council
\item[$^{\mathrm{b3}}$] 
 Supported by the Italian National Institute for Nuclear Physics (INFN)
\item[$^{\mathrm{b4}}$] 
 Supported by the German Federal Ministry for Education and Research (BMBF), under contract No. 05 H09PDF
\item[$^{\mathrm{b5}}$] 
 Supported by the Romanian National Authority for Scientific Research under the contract PN 09370101
\item[$^{\mathrm{b6}}$] 
 Supported by the Bundesministerium f\"ur Bildung und Forschung, FRG, under contract numbers 05H09GUF, 05H09VHC, 05H09VHF, 05H16PEA
\item[$^{\mathrm{b7}}$] 
 Supported by the SFB 676 of the Deutsche Forschungsgemeinschaft (DFG)
\item[$^{\mathrm{b8}}$] 
 Supported by VEGA SR grant no. 2/7062/ 27
\item[$^{\mathrm{b9}}$] 
 Supported by the National Science Centre under contract No. DEC-2012/06/M/ST2/00428
\item[$^{\mathrm{b10}}$] 
 Supported by HIR grant UM.C/625/1/HIR/149 and UMRG grants RU006-2013, RP012A-13AFR and RP012B-13AFR from Universiti Malaya, and ERGS grant ER004-2012A from the Ministry of Education, Malaysia
\item[$^{\mathrm{b11}}$] 
 Supported by CONACYT, M\'exico, grant 48778-F
\item[$^{\mathrm{b12}}$] 
 Russian Foundation for Basic Research (RFBR), grant no 1329.2008.2 and Rosatom
\item[$^{\mathrm{b13}}$] 
 Supported by RF Presidential grant N 3042.2014.2 for the Leading Scientific Schools and by the Russian Ministry of Education and Science through its grant for Scientific Research on High Energy Physics
\item[$^{\mathrm{b14}}$] 
 Supported by the Natural Sciences and Engineering Research Council of Canada (NSERC)
\item[$^{\mathrm{b15}}$] 
 Partially Supported by Polish Ministry of Science and Higher Education, grant DPN/N168/DESY/2009
\item[$^{\mathrm{b16}}$] 
 Partially Supported by Ministry of Science of Montenegro, no. 05-1/3-3352
\item[$^{\mathrm{b17}}$] 
 Supported by the Ministry of Education of the Czech Republic under the project INGO-LG14033
\item[$^{\mathrm{b18}}$] 
 Supported by the Israel Science Foundation
\item[$^{\mathrm{b19}}$] 
 Supported by the Japanese Ministry of Education, Culture, Sports, Science and Technology (MEXT) and its grants for Scientific Research
\item[$^{\mathrm{b20}}$] 
 Supported by the Swiss National Science Foundation
\item[$^{\mathrm{b21}}$] 
 Supported by DESY, Germany
\item[$^{\mathrm{b22}}$] 
 Also funded by Max Planck Institute for Physics, Munich, Germany, now at Sant Longowal Institute of Engineering and Technology, Longowal, Punjab, India
\item[$^{\mathrm{b23}}$] 
 Supported by the Initiative and Networking Fund of the Helmholtz Association (HGF) under the contract VH-NG-401 and S0-072
\item[$^{\mathrm{b24}}$] 
 Supported by the Alexander von Humboldt Foundation
\item[$^{\mathrm{b25}}$] 
 Also funded by Max Planck Institute for Physics, Munich, Germany, now at Sri Guru Granth Sahib World University, Fatehgarh Sahib, India
\item[$^{\mathrm{b26}}$] 
 Partially supported by the Polish National Science Centre projects DEC-2011/01/B/ST2/03643 and DEC-2011/03/B/ST2/00220
\end{description}}

\newpage
\section{Introduction}
\label{sect:introduction}

Measurements of open charm production in deep-inelastic electron\footnote{In 
this paper, `electron' is used to denote both electron and positron if not 
stated otherwise. }--proton scattering (DIS) at HERA provide important input 
for stringent tests of the theory of strong interactions, quantum 
chromodynamics ({QCD}).
Previous measurements \cite{h196,zeusdstar97,h1gluon,zd97,h1f2c,zd00,h1dmesons,h1vertex05,h1ltt_hera1,h1dstar_hera1,zeusdmesons,zd0dp,zmu,h1ltt_hera2,h1dstarhighQ2,zeusdpluslambda,h1cbjets,h1dstar_hera2,zeusdplus_hera2,zeusdstar_hera2} 
have demonstrated that charm quarks are predominantly produced by the 
boson--gluon-fusion process, $\gamma g\rightarrow c\overline{c}$, 
whereby charm production becomes 
sensitive to the gluon distribution in the proton.
Measurements have been obtained both from the HERA-I (1992--2000) and 
HERA-II (2003--2007) data-taking periods. 

Different techniques have been 
applied at HERA
to measure open-charm production 
in DIS.
The full reconstruction of $D$ or $D^{*\pm}$ mesons~\cite{
h196,zeusdstar97,h1gluon,zd97,h1f2c,zd00,h1dstar_hera1,zeusdmesons,zd0dp,h1dstarhighQ2,zeusdpluslambda,h1dstar_hera2, zeusdstar_hera2}, the long lifetime of heavy 
flavoured hadrons \cite{h1dmesons,h1vertex05,h1ltt_hera1,zd0dp,h1ltt_hera2,
h1cbjets,zeusdplus_hera2} 
or their semi-leptonic decays~\cite{zmu} are exploited. 
After extrapolation from the visible to the full phase space, 
most of these data have already been combined \cite{HERAcharmcomb} 
at the level of the reduced cross-sections 
and have provided a consistent determination of the charm contribution 
to the proton structure functions, 
a measurement of the charm-quark mass $m_c(m_c)$ and improved 
predictions for $W$- and $Z$-production cross sections at the LHC.
However, the extrapolation procedure requires theoretical assumptions, which 
lead to theoretical uncertainties comparable in size
to the experimental uncertainties~\cite{HERAcharmcomb}. 
Moreover, this combination was restricted to inclusive DIS variables only,
such as the photon virtuality, $Q^2$, and the inelasticity, $y$.
Alternatively, the measured cross sections can be combined directly in the
visible phase space. 
In this case, dependences on the theoretical input 
are minimised and the charm production mechanism can be explored 
in terms of other variables.
Such a combination, however, is possible only for 
data with the same final state, covering a common visible phase space.
The recently published differential cross-section measurements by H1 \cite{h1dstar_hera2,h1dstarhighQ2} and ZEUS \cite{zeusdstar_hera2} for inclusive $D^{*\pm}$-meson production fulfil this requirement. 
The analysis of fully reconstructed $D^{*\pm}$ mesons also offers the best signal-to-background ratio and small statistical uncertainties. 

In this paper, visible $D^{*\pm}$-production cross sections 
\cite{h1dstar_hera2,h1dstarhighQ2,zeusdstar_hera2,zd00} 
at the centre-of-mass energy $\sqrt{s}=318$~GeV are combined
such that one consistent HERA data set is obtained and compared directly 
to differential next-to-leading-order (NLO) QCD predictions.
The combination is based on the procedure described 
elsewhere \cite{HERAcharmcomb,glazov,H1comb,DIScomb}, 
accounting for all correlations in the uncertainties. 
This yields a significant 
reduction of the overall uncertainties of the measurements. 
The possibility to describe all measurements both in shape and normalisation
with a single set of theory parameter values is also investigated and 
interpreted in terms of future theory improvements.

The paper is organised as follows. In Section~\ref{sect:theory} the theoretical framework is briefly introduced that is used for applying phase-space corrections 
to the input data sets prior to combination and for providing NLO QCD predictions to be compared to the data. 
The data samples used for the combination are detailed in Section~\ref{sect:samples} and the combination procedure is described in Section~\ref{sect:combine}. 
The combined single- and double-differential cross sections are presented in Section~\ref{sect:singlediff} 
together with a comparison of NLO QCD predictions to the data.

\section{Theoretical predictions}
\label{sect:theory}

The massive fixed-flavour-number scheme (FFNS)~\cite{ffns} is used for theoretical predictions, since it is the only scheme for which fully differential NLO calculations \cite{hvqdis} are available.
The cross-section predictions for $D^{*\pm}$ production presented in this paper
are obtained 
using the HVQDIS program \cite{hvqdis} which provides NLO QCD  ($O(\alpha_s^2)$) calculations in the 3-flavour FFNS for charm and beauty production in DIS. 
These predictions are used both 
for small phase-space corrections of the data, due to slightly 
different binning schemes and kinematic cuts, and 
for comparison to data.

The following parameters are used in the calculations and 
are varied within certain limits to estimate the uncertainties associated with the predictions:
\begin{itemize}
 \item The {\bf renormalisation and factorisation scales}
are taken as $\mu_r=\mu_f=\sqrt{Q^2+4m_c^2}$. The scales are varied simultaneously up or down by
  a factor of two for the phase-space corrections 
 where only the shape of the differential cross sections is relevant. 
For absolute predictions, the scales are changed independently to $0.5$ and $2$ times their nominal value. 
 \item The {\bf pole mass of the charm quark} is set to  $m_c=1.50 \pm 0.15$ GeV. This variation also affects 
the values of the renormalisation and factorisation scales. 
 \item For the {\bf strong coupling constant} the value $\alpha_s^{n_f=3}(M_Z) = 0.105 \pm 0.002$ is chosen \cite{HERAcharmcomb} which corresponds to
 $\alpha_s^{n_f=5}(M_Z) = 0.116 \pm 0.002$.
\item The {\bf proton parton density functions (PDFs)} are described by a 
series of {FFNS} variants of the HERAPDF1.0 set~\cite{DIScomb} 
at NLO determined within the HERAFitter~\cite{herafitter} framework, similar to those used 
in the charm combination paper
\cite{HERAcharmcomb}. 
Charm measurements were not included in the determination of these PDF sets.
For all parameter settings used here, the corresponding PDF set is used.
By default, the scales for the charm contribution to the inclusive 
data in the PDF determination were chosen to be consistent with the factorisation scale 
used in HVQDIS, while the 
renormalisation scale in HVQDIS was decoupled from the scale used in the PDF extraction, 
except
in the cases where the factorisation and renormalisation scales were varied 
simultaneously.
As a cross check, the renormalisation scales for both heavy- and light-quark
contributions are varied simultaneously in HVQDIS and in the PDF 
determination, keeping the factorisation scales fixed. 
The result lies well within the 
quoted uncertainties.
The cross sections are also evaluated with 3-flavour NLO versions of the ABM~\cite{abm11} and MSTW~\cite{mstw08f3} PDF sets.
The differences are found to be negligible compared to those from varying 
other parameters, such that no attempt for coverage of all possible PDFs 
is made.

\end{itemize}

The NLO calculation performed by the HVQDIS program yields 
differential cross sections for charm quarks. These predictions are converted to $D^{*\pm}$-meson cross sections by applying the fragmentation model described in a previous
publication~\cite{HERAcharmcomb}. 
This model is based on the fragmentation function of Kartvelishvili et
al.~\cite{Kartvelishvili:1977pi} which provides a probability density function for the fraction of the charm-quark momentum transferred to the $D^{*\pm}$ meson. 
The function is controlled by a single fragmentation parameter, $\alpha_K$.
Different values of $\alpha_K$ \cite{HERAcharmcomb} 
are used for different regions of the invariant mass, $\hat s$, of
the photon--parton centre-of-mass system. 
The boundary $\hat s_1 = 70 \pm 40$~GeV$^2$ between
the first two regions is one of the parameter variations.
The boundary $\hat s_2 = 324$~GeV$^2$ 
between the second and third region remains fixed.
The model also implements a transverse-fragmentation
component by assigning to the $D^{*\pm}$ meson a transverse momentum, $k_T$,
with respect  to the charm-quark direction~\cite{HERAcharmcomb}.
The following parameters are used in the calculations 
together with the 
corresponding variations 
for estimating the uncertainties of the NLO predictions
related to fragmentation:

\begin{itemize}
\item The {\bf fragmentation parameter $\pmb{\alpha_K}$, the bin boundary $\pmb{\hat s_1}$ and the average $\pmb{k_T}$} are
varied according to a 
prescription described elsewhere~\cite{HERAcharmcomb}.
\item The {\bf fraction of charm quarks hadronising into $\pmb{D^{*+}}$ mesons} is set to\\ 
$f(c\rightarrow D^{*+})=f(\bar c\rightarrow D^{*-})=0.2287 \pm 0.0056$~\cite{Lohrmann:2011ce}.
\end{itemize}

The small beauty contribution to the $D^{*\pm}$ signal needs a detailed treatment of the
$B$ hadron decay to $D^{*\pm}$ mesons and is therefore obtained from NLO QCD predictions for beauty hadrons 
convoluted with decay tables to $D^{*\pm}$ mesons and decay kinematics obtained 
from EvtGen \cite{EvtGen}. 
The parameters for the calculations and the uncertainties are:
\begin{itemize}
\item The {\bf renormalisation and factorisation scales} 
$\mu_r=\mu_f=\sqrt{Q^2+4m_b^2}$
are varied in the same way as described above for charm. The variations are applied simultaneously for the calculation of the charm and beauty cross-section uncertainties.
\item The {\bf pole mass of the beauty quark} is set to $m_b=4.75 \pm 0.25$ GeV.
\item The {\bf  fragmentation model for beauty quarks}  is based on 
the Peterson et al. \cite{Peterson} parametrisation 
using $\epsilon_{b}=0.0035 \pm 0.0020$~\cite{epsilonb}.
\item The {\bf fraction of beauty hadrons decaying into $\pmb{D^{*\pm}}$ mesons} is set to\\
$f(b\rightarrow D^{*\pm})=0.173 \pm 0.020$~\cite{pdg14}.
\item The {\bf proton structure} is described by the same PDF set 
(3-flavour scheme) used for the charm cross-section predictions.
\end{itemize}

The total theoretical uncertainties are obtained by adding all individual contributions 
in quadrature.

\section{Data samples for cross-section combinations}
\label{sect:samples}
The H1\cite{h1.1,h1.2,h1.3} and ZEUS~\cite{zeus} detectors 
were general purpose
instruments which consisted of tracking systems surrounded by
electromagnetic and hadronic calorimeters and muon detectors. 
The most important detector components for the measurements combined in this paper are the central tracking detectors (CTD) operated inside solenoidal magnetic fields of $1.16$~T~(H1) and $1.43$ T (ZEUS) and the electromagnetic sections of the calorimeters.
The CTD  of H1~\cite{h1.2}  (ZEUS \cite{ZEUSCTD}) measured charged particle trajectories in the polar angular range\footnote{In both experiments a right-handed coordinate system is employed with the $Z$ axis pointing in the nominal proton-beam direction, referred to as ``forward direction'', and the $X$ axis pointing towards the centre of HERA. The origin of the coordinate system is defined by the nominal interaction point in the case of H1 and by the centre of the CTD in the case of ZEUS.
}
 of $15^\circ < \Theta< 165(164)^\circ$. 
In both detectors the CTDs
were 
complemented with high-resolution silicon vertex detectors: a system of three silicon detectors for H1, consisting of the Backward Silicon Tracker~\cite{H1bst},  the Central Silicon Tracker \cite{H1cst} and the Forward Silicon Tracker~\cite{H1fst}, and the Micro Vertex Detector \cite{ZEUSmvd} for ZEUS. 
For charged particles passing through all active layers of the silicon vertex detectors and CTDs, transverse-momentum resolutions of 
$\sigma(p_T)/p_T\simeq 0.002p_T/ \oplus 0.015$~(H1) and $\sigma(p_T)/p_T \simeq 0.0029p_T/\oplus 0.0081 \oplus 0.0012/p_T$~(ZEUS), with $p_T$ in units of GeV, have been achieved. 

Each of the central tracking detectors was enclosed by a set of calorimeters comprising an inner electromagnetic and an outer hadronic section. 
The H1 calorimeter system consisted of the Liquid Argon calorimeter (LAr)~\cite{H1Lar} 
and the backward lead--scintillator calorimeter (SpaCal)~\cite{h1.3} while the ZEUS detector was equipped with a compensating  uranium--scintillator calorimeter (CAL)~\cite{ZEUSCAL}.
Most important for the analyses combined in this paper is the electromagnetic part of the calorimeters which is used to identify and measure the scattered electron. Electromagnetic energy resolutions 
$\sigma(E)/E$ of $0.11/\sqrt{E}$ (LAr)~\cite{H1Lartest}, $0.07/\sqrt{E}$ (SpaCal)~\cite{H1spacaltest} and  $0.18/\sqrt{E}$ (CAL), with $E$ in units of GeV, were achieved.

The Bethe--Heitler process, $ep\rightarrow e\gamma p$, is used by both experiments to determine the luminosity. Photons originating from this reaction were detected by photon taggers at about $100$~m downstream of the electron beam line.  The integrated luminosities are known with a precision of $3.2$\% for the H1 measurements~\cite{h1dstar_hera2,h1dstarhighQ2} and of about $2$\% for the ZEUS measurements~\cite{zeusdstar_hera2,zd00,ZEUSlumi2}.


\begin{table}[t]
\begin{center}
\tabcolsep1.1mm
\renewcommand*{\arraystretch}{1.2}
\begin{tabular}{|r|l|r|l|l|l|l|r|}
\hline
\multicolumn{3}{|c|}{\multirow{3}{*}{Data set}} & \multicolumn{4}{|c|}{Kinematic range}      &   \\ \cline{4-7}
\multicolumn{3}{|c|}{}                         & \qquad $Q^2$        & \qquad $y$ & \ \ $p_{T}(D^*)$ & \quad $\eta(D^*)$ & ${\cal L}$\ \ \\ 
\multicolumn{3}{|c|}{}                         & \ \ (\gev$^{2}$) &     & \ \ (\gev)     &             & ($\text{pb}^{\text{-1}}$)            \\ \hline
I & H1 $D^{*\pm}$ HERA-II (medium $Q^2$) & \cite{h1dstar_hera2} & $~~~~5:~~100$ & $0.02:0.70$ & $~~>1.5$ & $-1.5:1.5$ & $348$ \\ 
II & H1 $D^{*\pm}$ HERA-II (high $Q^2$) & \cite{h1dstarhighQ2} & $100:1000$ & $0.02:0.70$ & $~~>1.5$ & $-1.5:1.5$ & $351$ \\ 
III & ZEUS $D^{*\pm}$ HERA-II & \cite{zeusdstar_hera2} & $~~~~5:1000$ & $0.02:0.70$ & $1.5:20.0$ & $-1.5:1.5$ & $363$ \\ 
IV & ZEUS $D^{*\pm}$ 98-00 & \cite{zd00} & $~1.5:1000$ & $0.02:0.70$ & $1.5:15.0$ & $-1.5:1.5$ & $~~82$ \\ \hline
\end{tabular}
\end{center}
\caption{Data sets used in the combination. For each data set the respective kinematic range 
and the integrated luminosity, ${\cal L}$, are given.}
\label{tab:input}
\end{table}

Combinations are made for single- and double-differential cross sections.
In Table~\ref{tab:input} the datasets\footnote{Of the two sets of measurements in \cite{h1dstar_hera2}, that
compatible with the above cuts is chosen.
} used for these combinations are listed together with their visible phase-space regions and integrated luminosities.
The datasets I--III are used to determine single-differential combined cross sections 
 as a function of the $D^{*\pm}$ meson's 
transverse momentum, $p_T(D^{*})$,  pseudorapidity, 
$\eta(D^{*})$, 
and inelasticity, 
$z(D^{*})= (E(D^{*})-p_Z(D^{*}))/(2E_e y)$, measured in the laboratory frame,
and of $Q^2$ and $y$.
The variables $E(D^{*})$, $p_Z(D^{*})$ and $E_e$ denote the energy of the $D^{*\pm}$ meson, the $Z$ component of the momentum of the $D^{*\pm}$ meson and the incoming electron energy, respectively.
Owing to beam-line modifications related to the HERA-II high-luminosity running \cite{Upgrade} the visible phase space of these cross sections at HERA-II is restricted to $Q^2 > 5$~GeV$^2$, which prevents a combination with earlier $D^{*\pm}$ cross-section measurements for which the phase space extends 
down to $Q^2=1.5$~GeV$^2$.

In the case of the double-differential cross section, ${\rm d}^2\sigma/{\rm d}y{\rm d}Q^2$, 
the kinematic range can be extended to lower $Q^2$ using HERA-I measurements~\cite{h1dstar_hera1,zd97,zd00}. 
In order to minimise the use of correction factors derived from theoretical calculations, 
the binning scheme of such measurements has to be similar to that used for the HERA-II data. One of the 
HERA-I measurements, set~IV of Table \ref{tab:input}, satisfies this requirement and is therefore included in the combination of this double-differential cross section. 
The visible phase spaces of the combined single- and double-differential cross sections are summarised in Table~\ref{Tab:vps}.

The measurements 
to be combined for the single- and double-differential cross sections 
are already corrected 
to the Born level with a running fine-structure constant $\alpha$ and 
include both the charm and beauty contributions to $D^{*\pm}$ production. 
The total expected beauty contribution is small, varying from $\sim 1\%$ at 
the lowest $Q^2$ to $\sim 7\%$ at the highest $Q^2$. 
The cross sections measured previously~\cite{zd00,h1dstar_hera2,h1dstarhighQ2} 
are here corrected to the 
PDG value \cite{pdg14} of the $D^0$ branching ratio.


\begin{table}[t]
\renewcommand{\arraystretch}{1.3}
\begin{center}
\begin{tabular}{|lr|ccc|}
\hline
&&single&&double\cr
\multicolumn{2}{|c}{Range in}&\multicolumn{3}{|c|}{differential cross section}\cr
\hline
$Q^2$ &(GeV$^2$)&$5-1000$&\hspace*{.5cm} &$1.5-1000$\cr
$y$&&\multicolumn{3}{|c|}{$0.02-0.7$}\cr
$p_T(D^*)$& (GeV)&\multicolumn{3}{|c|}{$> 1.5$}\cr
$|\eta(D^*)|$&&\multicolumn{3}{|c|}{$< 1.5$}\cr
\hline
\end{tabular}
\end{center}
\caption{Visible phase space of the combined cross sections.} 
\label{Tab:vps}
\end{table}

\subsection{ Treatment of data sets for single-differential cross sections}

In order to make the input data sets compatible with the phase space 
quoted in Table \ref{Tab:vps} and 
with each other, the following corrections are applied before the combination:
\begin{itemize}
\item The H1 collaboration has published measurements of $D^{*\pm}$ 
cross sections separately for $5$~GeV$^2<Q^2<100$~GeV$^2$ 
(set~I)
 and for $100$~GeV$^2<Q^2<1000$~GeV$^2$ 
(set~II). 
Due to the limited statistics at high $Q^2$, 
a coarser binning in $p_T(D^*)$, $\eta(D^*)$, $z(D^*)$ and $y$ was used in 
set~II 
compared to set~I. 
Therefore the cross section in a bin $i$ of a given observable integrated in the range $5$~GeV$^2<Q^2<1000$~GeV$^2$ is calculated according to
\begin{eqnarray}
\lefteqn{\sigma_i(5<Q^2/\text{GeV}^2<1000)=\sigma_i(5<Q^2/\text{GeV}^2<100)}\\\nonumber
&&{}+\sigma_i^{\rm NLO}(100<Q^2/\text{GeV}^2<1000)\cdot\frac{\sigma_{\rm int}(100<Q^2/\text{GeV}^2<1000)}{\sigma_{\rm int}^{\rm NLO}(100<Q^2/\text{GeV}^2<1000)}.
\end{eqnarray} 
Here $\sigma_{\rm int}$ denotes the integrated visible cross section and $\sigma^{\rm NLO}$ stands for the NLO predictions 
obtained from HVQDIS. In this calculation both the experimental uncertainties of the visible cross section at high $Q^2$ and the theoretical uncertainties as described in Section~\ref{sect:theory} are included. 
The contribution from the region $100$~GeV$^2<Q^2<1000$~GeV$^2$ to the full $Q^2$ range amounts to $4$\% on average and reaches up to $50$\% at highest $p_T(D^*)$.

\item
The bin boundaries used for the differential cross section as a function of $Q^2$ differ between sets~I, II and set III. At low $Q^2$ this is solved by combining the cross-section measurements of the first two bins of set~I into a single bin. For $Q^2>100$~GeV$^2$ no consistent binning scheme could be defined directly from the single-differential cross-section  measurements. 
However, the measurements of the double-differential cross sections ${\rm d}^2\sigma/{\rm d}Q^2{\rm d}y$ have been performed in a common binning scheme. By integrating these cross sections in $y$, single-differential cross sections in $Q^2$ are obtained also for  $Q^2>100$~GeV$^2$ which can be used directly in the combination.

\item

The cross-section measurements in set~III are restricted to $p_T(D^*)<20$~GeV while there is no such limitation in the phase space of the combination. Therefore these cross sections are corrected for the contribution from the range $p_T(D^*)>20$~GeV using HVQDIS. This correction is found to be less than $0.1$\%. 
\end{itemize}

\subsection{ Treatment of data sets for double-differential cross sections}

Since the restriction to the same phase space in $Q^2$ does not apply for the combination of the double-differential cross sections in $Q^2$ and $y$, 
the HERA-I measurement, set~IV, is also included in the combination.
This allows an extension of the kinematic range down to $Q^2>1.5~{\rm GeV}^2$.
The $p_T(D^*)$ ranges of the measurements of sets III and IV are corrected in the same way as for the single-differential cross sections.

To make the binning scheme of the HERA-I measurement compatible with that used for the HERA-II datasets, the binning for all datasets is revised. 
Cross sections in the new bins are obtained from the original bins using the shape of the HVQDIS predictions as described in Section~\ref{sect:theory}.
The new binning is given in Section~\ref{sect:singlediff} (Table \ref{tab:result_q2y}). 
Bins are kept only if they satisfied both of the following criteria:
\begin{itemize}
\item The predicted fraction of the cross section of the original bin contained in the kinematical overlap region in $Q^2$ and $y$ between the original and corrected bins is greater than 50\% (in most bins it is greater than 90\%).
\item The theoretical uncertainty from the correction procedure is obtained by evaluating all uncertainties discussed in Section~\ref{sect:theory} and adding them in quadrature. 
The ratio of the theoretical uncertainty to the uncorrelated experimental uncertainty is required to be less than 30\%. 
\end{itemize}
This procedure ensures that the effect of the theoretical uncertainties on the combined data points is small. 
Most of the HERA-II bins are left unmodified; all of them satisfied the criteria and are kept. Out of the 31 original HERA-I bins, 26 bins satisfy the criteria and are kept. The data points removed from the combination mainly correspond to the low-$y$ region where larger bins were used for the HERA-I data.

\section{Combination method}
\label{sect:combine}
The combination of the data sets uses the $\chi^2$ minimisation method 
developed for the 
combination of inclusive DIS cross sections~\cite{glazov,DIScomb}, as 
implemented in the HERAverager program~\cite{HERAAverager}.
For an individual dataset $e$ the contribution to the $\chi^2$ function is defined as
\begin{equation}
\chi^2_{{\rm exp},e}(m^i,b_j) = 
 \sum_i
 \frac{\left(m^i - \sum_j \gamma^{i,e}_j m^i b_j  - {\mu^{i,e}} \right)^2}
      { \left(\delta_{i,e,{\rm stat}}\,\mu^{i,e}\right)^2+
        \left(\delta_{i,e,{\rm uncor}}\,m^i\right)^2}\,.
\label{eq:ave}
\end{equation}
Here  ${\mu^{i,e}}$ is the measured value of the cross section in bin $i$
and $\gamma^{i,e}_j $, $\delta_{i,e,{\rm stat}} $ and 
$\delta_{i,e,{\rm uncor}}$ are the relative
correlated systematic, relative statistical and relative uncorrelated systematic uncertainties,
respectively, from the original measurements.
The quantities $m^i$ express the values 
of the expected combined cross section for each bin $i$ and the quantities $b_j$ 
express the shifts 
of the correlated systematic-uncertainty sources $j$, in units of the standard deviation. 
Several data sets providing a number of measurements (index $e$) 
are represented
by a total $\chi^2$ function,
which is built from the sum of the $\chi^2_{{\rm exp},e}$ functions of all data sets
\begin{equation}
\chi^2_{\rm tot}(m^i,b_j) = \sum_e \chi^2_{{\rm exp},e}(m^i,b_j) + \sum_j b^2_j\,. \label{eq:tot}
\end{equation}
The combined cross sections $m^i$
are 
obtained by the minimisation of
$\chi^2_{\rm tot}$
with respect to $m^i$ and $b_j$.

The averaging procedure also provides the covariance matrix of the
$m^i$ and the uncertainties of the $b^j$ at the minimum.
%
The $b_j$ at the minimum and their uncertainties 
are referred to as ``shift'' and ``reduction'', respectively.
The covariances $V$ of the $m^i$ are given in the form
$V=V_{\rm uncor}+ \sum_k V_{\rm sys}^k$~\cite{H1comb}.
%
The matrix $V_{\rm uncor}$ is diagonal.
Its diagonal elements correspond to the covariances obtained in a 
weighted average performed in the absence of any correlated systematic
uncertainties.
The covariance matrix contributions $V_{\rm sys}^k$ correspond to
correlated systematic uncertainties on the averaged cross sections,
such that the elements of a matrix $V_{\rm sys}^k$ are obtained as
$(V_{\rm sys}^k)_{ij}=\delta^{{\rm sys},k}_i\delta^{{\rm sys},k}_j$, given a vector
$\delta^{{\rm sys},k}$ of systematic uncertainties.
It is worth noting that, in this representation of the covariance
matrix, the number of correlated systematic sources is identical to
the number of correlated systematic sources in the input data sets.

In the present analysis, the correlated and uncorrelated systematic uncertainties are predominantly of multiplicative 
nature, i.e. they change proportionally to the central values. In equation~(\ref{eq:ave}) the 
multiplicative nature of these uncertainties is taken into account by multiplying the 
relative errors $\gamma^{i,e}_j$ and $\delta_{i,e,{\rm uncor}}$ by the 
cross-section expectation $m^i$.
In charm analyses the statistical uncertainty is mainly background dominated. 
Therefore it is treated as being independent of $m^i$. 
For the minimisation of $\chi^2_{\rm tot}$ an iterative procedure is used as described elsewhere~\cite{H1comb}.

The $55$ systematic uncertainties obtained from the original publications 
were examined for their correlations.
Within each data set, most of the systematic uncertainties are found to be 
point-to-point correlated, and are thus treated as fully correlated 
in the combination. 
In total there are $23$ correlated experimental systematic sources 
and $5$ theory-related uncertainty sources. 
A few are found to be uncorrelated and added in quadrature. 
For the combination of single-differential cross sections 
the uncorrelated uncertainties also include a theory-related uncertainty 
from the corrections discussed in Section \ref{sect:samples}, 
which varies between 0 and 10\% of the total uncertainty and is added in 
quadrature. 
Asymmetric systematic uncertainties were symmetrised to the larger deviation 
before performing the combination.
Except for the branching-ratio uncertainty, which was treated as correlated, 
all experimental systematic uncertainties were treated as independent between the 
H1 and ZEUS data sets. 
Since the distributions in $p_T(D^*)$, $\eta(D^*)$, $z(D^*)$, $Q^2$ and $y$ 
are not statistically independent, each distribution is combined separately.

\section{Combined cross sections}
\label{sect:singlediff}

The results of combining the HERA-II measurements 
\cite{h1dstar_hera2,h1dstarhighQ2,zeusdstar_hera2}
as a function of $p_T(D^*)$, $\eta(D^*)$, $z(D^*)$, $Q^2$ and $y$  are given 
in Tables~\ref{tab:result_pt} -- \ref{tab:result_y}, together with their uncorrelated and correlated uncertainties\footnote{
A detailed breakdown of correlated uncertainties can be 
found on \\
\url{http://www.desy.de/h1zeus/dstar2015/}.}.
The total uncertainties are obtained by adding the uncorrelated and correlated 
uncertainties in quadrature.

\begin{figure}[h]
\center
\epsfig{file=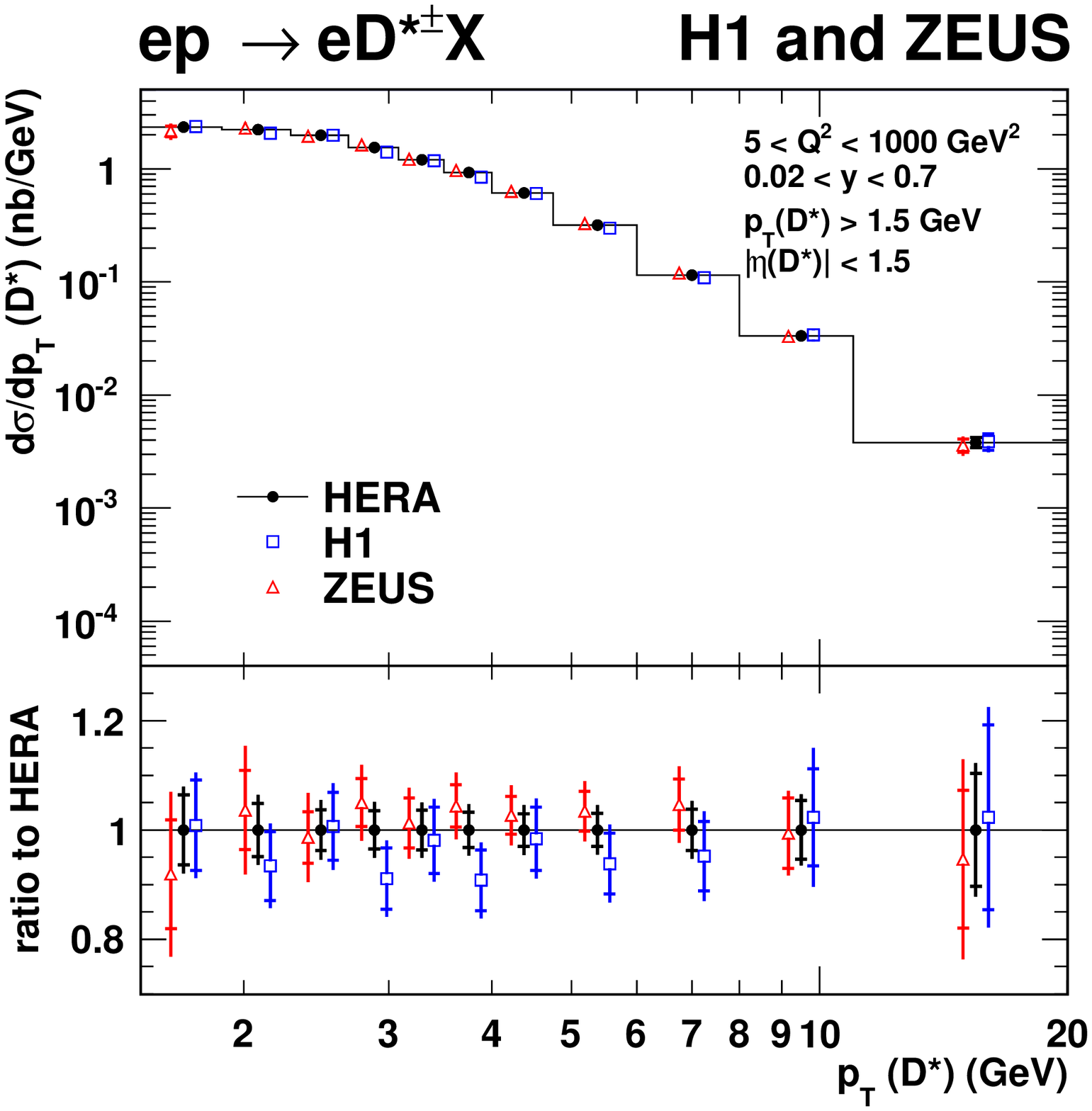,width=1.0\textwidth}
\caption{Differential $D^{*\pm}$-production cross section as a function of $p_T(D^{*})$. 
The open triangles 
and squares are the cross sections before combination, shown with a 
small horizontal offset for better visibility. 
The filled points are the combined cross sections. The inner 
error bars indicate the uncorrelated part of the uncertainties.
The outer error bars represent the total uncertainties.
The histogram indicates the binning used to calculate the cross sections.
The bottom part shows the ratio of these cross sections with respect to the 
central value of the combined cross sections.}
\label{fig:pt} 
\end{figure}

\begin{figure}[h]
\center
\epsfig{file=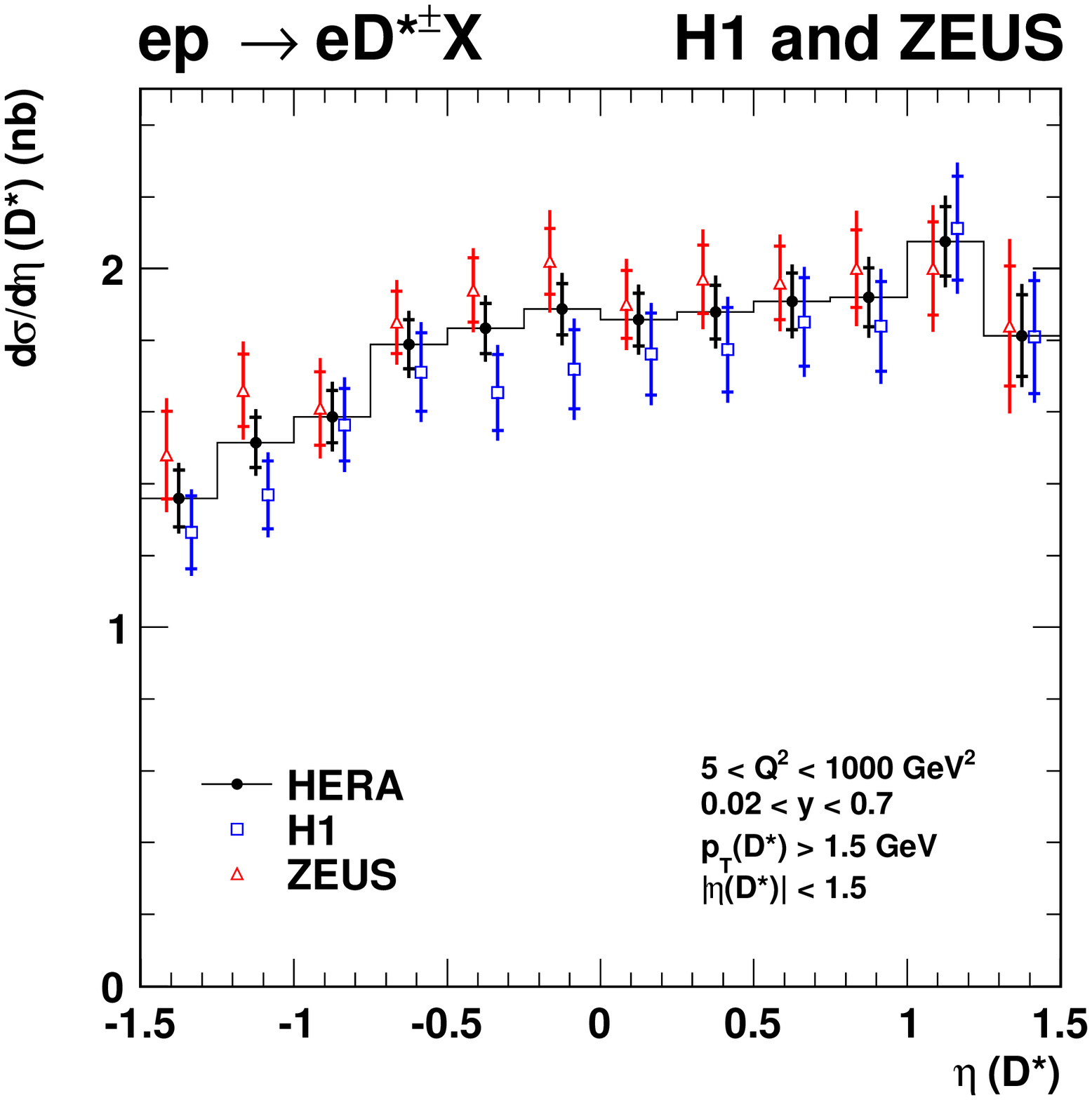,width=1.0\textwidth}
\caption{Differential $D^{*\pm}$-production cross section as a function of $\eta(D^{*})$. 
The open triangles 
and squares are the cross sections before combination, shown with a 
small horizontal offset for better visibility. 
The filled points are the combined cross sections. The inner 
error bars indicate the uncorrelated part of the uncertainties.
The outer error bars represent the total uncertainties.
The histogram indicates the binning used to calculate the cross sections.
}
\label{fig:eta} 
\end{figure}

\begin{figure}[h]
\center
\epsfig{file=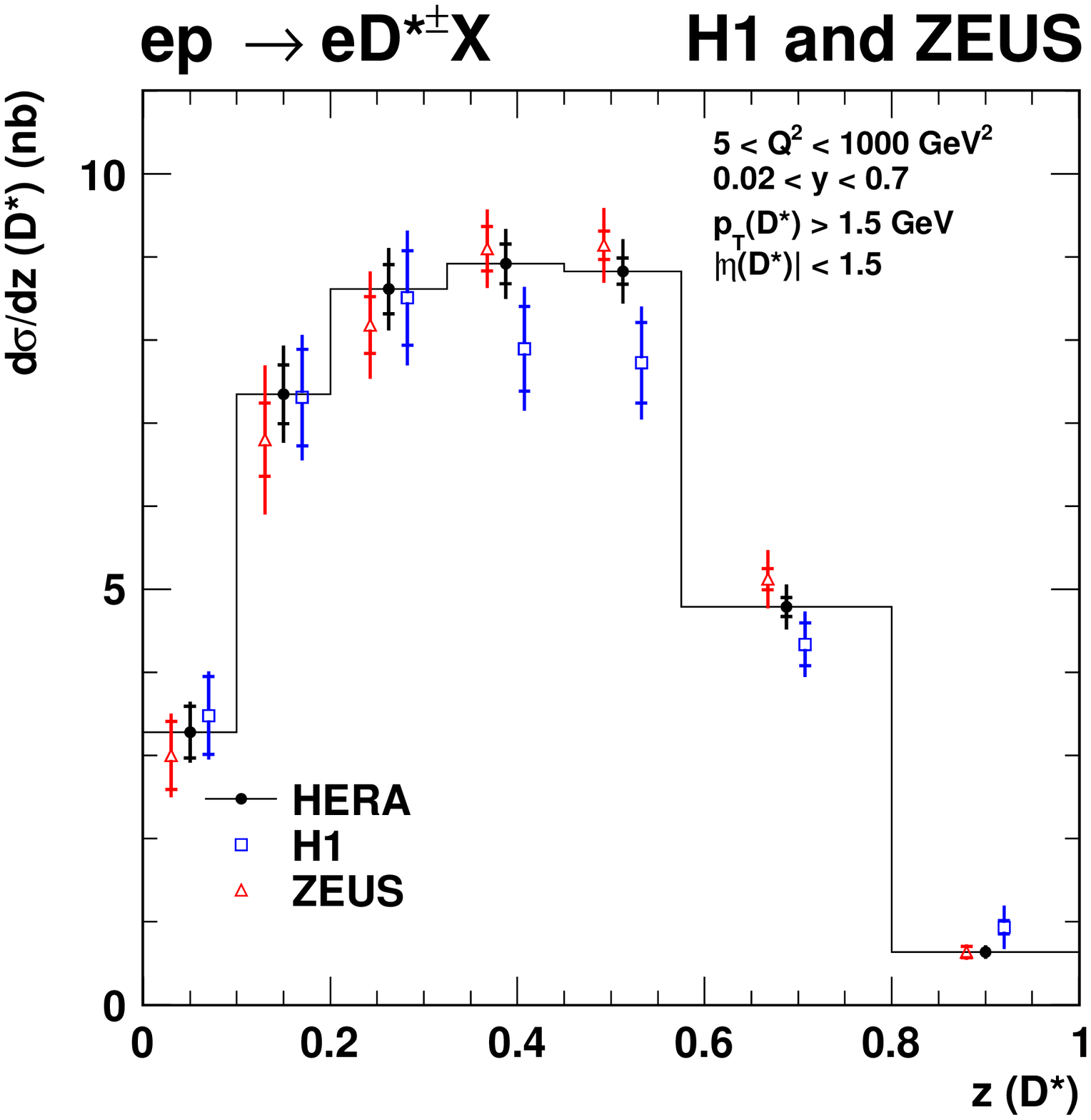,width=1.0\textwidth}
\caption{Differential $D^{*\pm}$-production cross section as a function of $z(D^{*})$. 
The open triangles 
and squares are the cross sections before combination, shown with a 
small horizontal offset for better visibility. 
The filled points are the combined cross sections. The inner 
error bars indicate the uncorrelated part of the uncertainties.
The outer error bars represent the total uncertainties.
The histogram indicates the binning used to calculate the cross sections.
}
\label{fig:z} 
\end{figure}

\begin{figure}[h]
\center
\epsfig{file=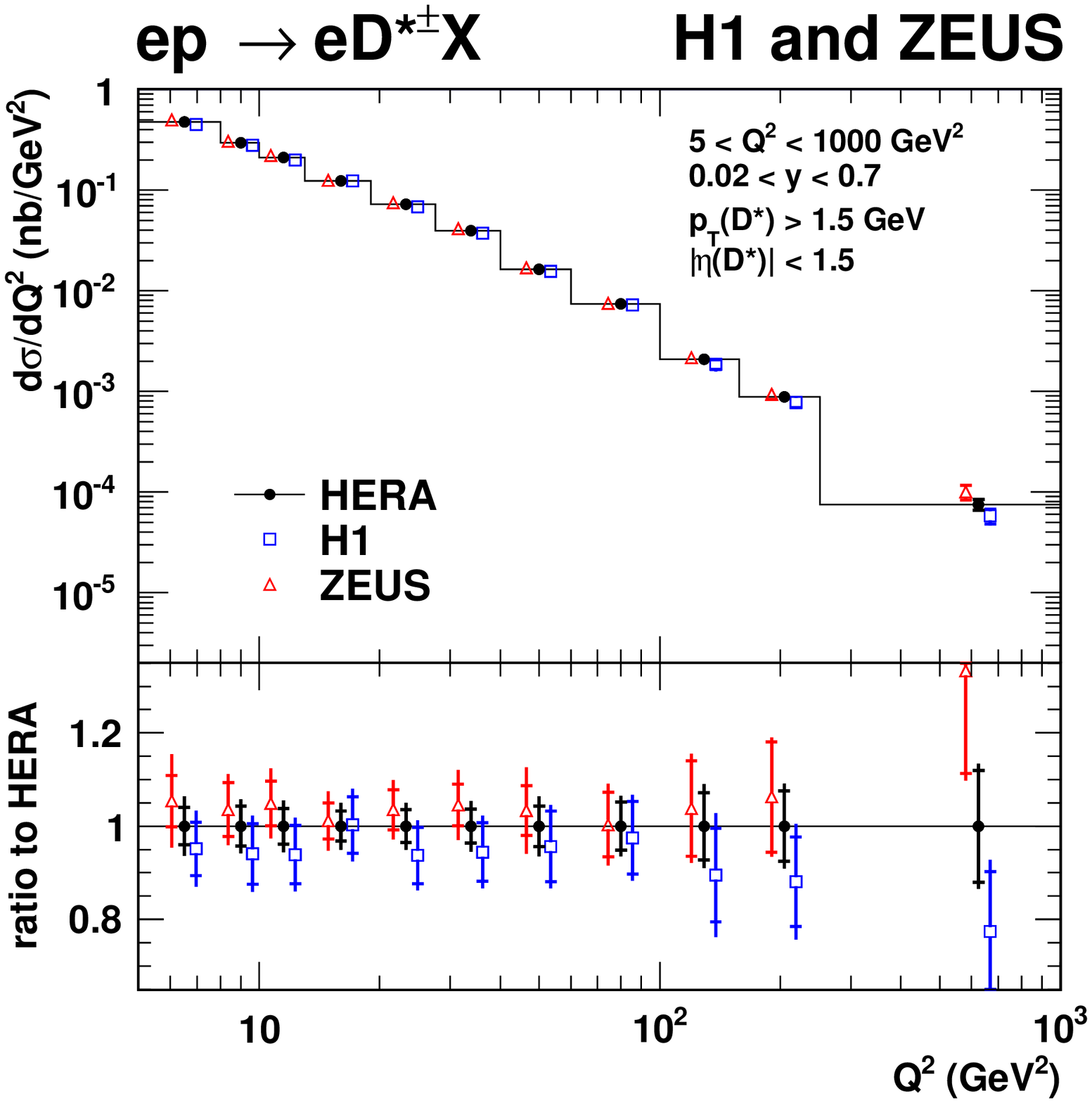,width=1.0\textwidth}
\caption{Differential $D^{*\pm}$-production cross section as a function of $Q^2$. 
The open triangles 
and squares are the cross sections before combination, shown with a 
small horizontal offset for better visibility. 
The filled points are the combined cross sections. The inner 
error bars indicate the uncorrelated part of the uncertainties.
The outer error bars represent the total uncertainties.
The histogram indicates the binning used to calculate the cross sections.
The bottom part shows the ratio of these cross sections with respect to the 
central value of the combined cross sections.}
\label{fig:q2} 
\end{figure}

\begin{figure}[h]
\center
\epsfig{file=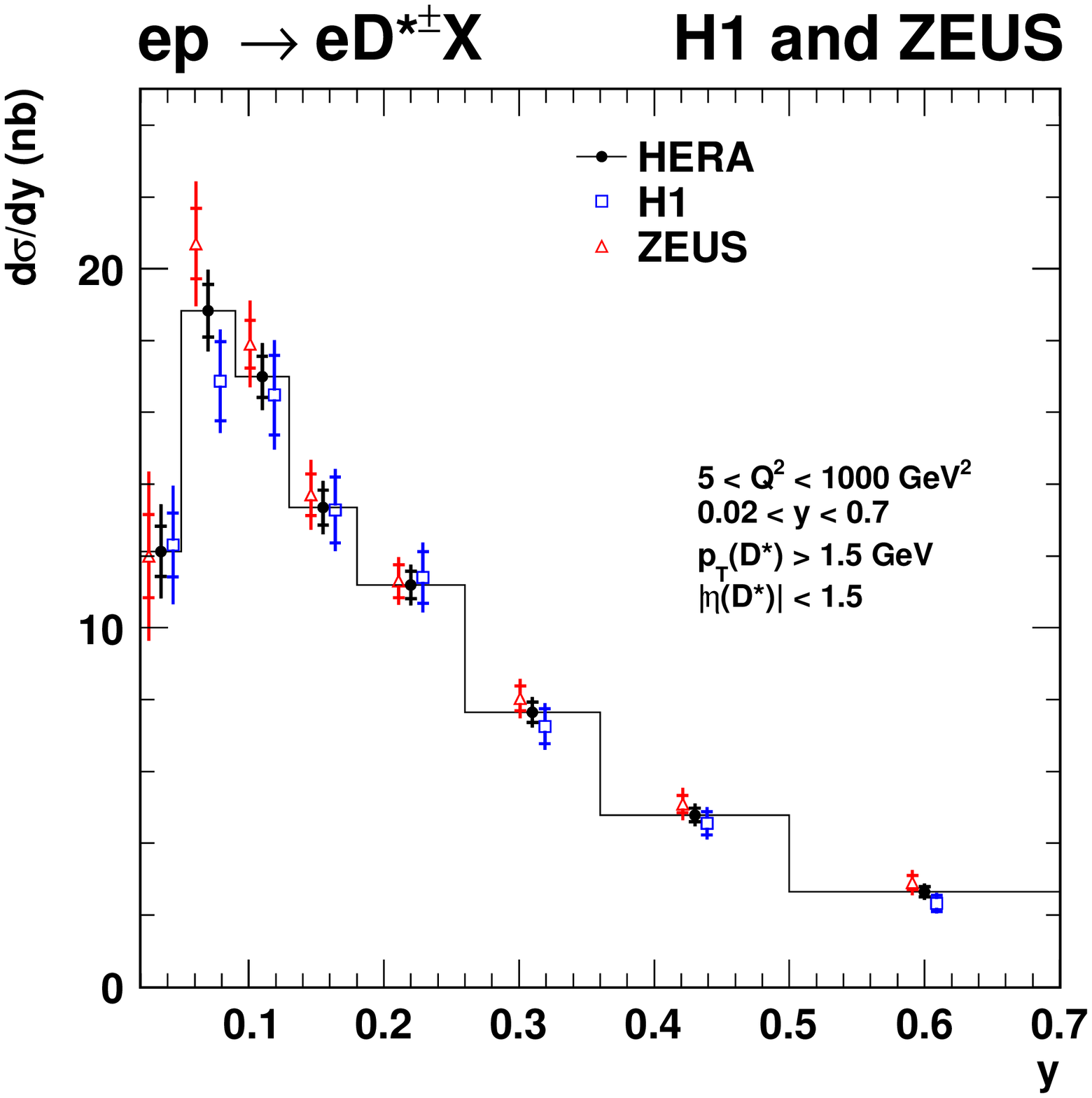,width=1.0\textwidth}
\caption{Differential $D^{*\pm}$-production cross section as a function of $y$. 
The open triangles 
and squares are the cross sections before combination, shown with a 
small horizontal offset for better visibility. 
The filled points are the combined cross sections. The inner 
error bars indicate the uncorrelated part of the uncertainties.
The outer error bars represent the total uncertainties.
The histogram indicates the binning used to calculate the cross sections.
}
\label{fig:y} 
\end{figure}

The individual data sets and the results of the combination are shown
in Figures~\ref{fig:pt} -- \ref{fig:y}.  
The consistency of the data sets as well as the reduction of the 
uncertainties are illustrated further by the insets at the bottom of 
Figures~\ref{fig:pt} and \ref{fig:q2}. 
The combinations in the different variables have a $\chi^2$ probability 
varying between 15\% and 87\%, i.e.\ the data sets are consistent.
The systematic shift between the two input data sets is covered by the 
respective correlated uncertainties.
The shifts and reductions of the correlated uncertainties are given in Table~\ref{tab:syst}. 
The improvement of the total correlated uncertainty is due to small reductions of many sources.
While the effective doubling of the statistics of the combined result reduces 
the uncorrelated uncertainties, 
the correlated uncertainties 
of the combined cross sections are reduced through 
cross-calibration effects between the two experiments. Typically, both effects 
contribute about equally to the reduction of the total uncertainty.

\begin{figure}[h]
\center
\epsfig{file=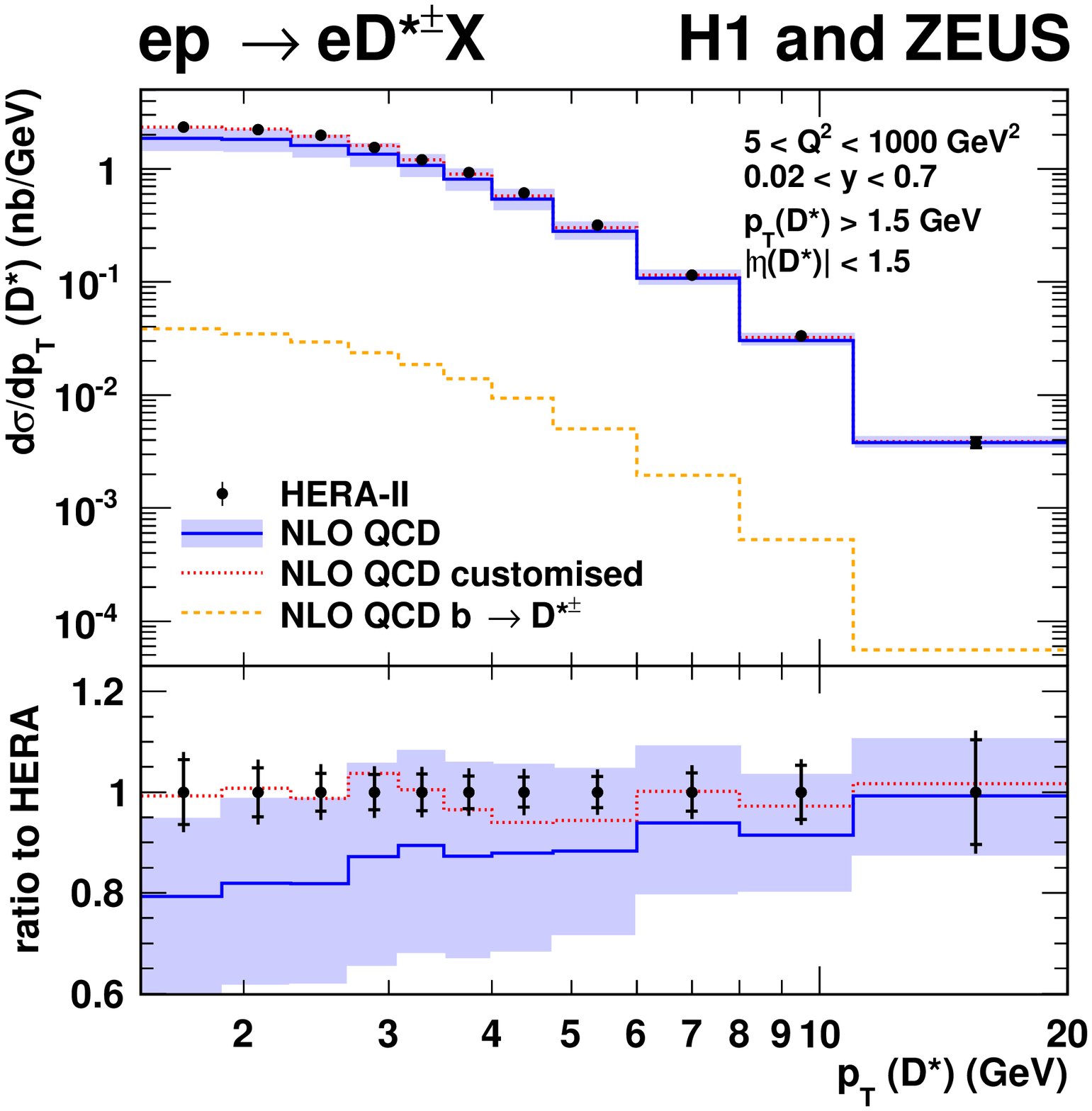,width=1.0\textwidth}
\caption{Differential $D^{*\pm}$-production cross section as a function of $p_T(D^{*})$.
The data points are the combined cross sections. The inner 
error bars indicate the uncorrelated part of the uncertainties.
The outer error bars represent the total uncertainties.
Also shown are the NLO predictions from HVQDIS (including the beauty 
contribution) and their uncertainty band.
A customised NLO calculation (dotted line, see text) is also shown.
The bottom part shows the ratio of these cross sections with respect to the 
central value of the combined cross sections.}
\label{fig:ptth} 
\end{figure}

\begin{figure}[h]
\center
\epsfig{file=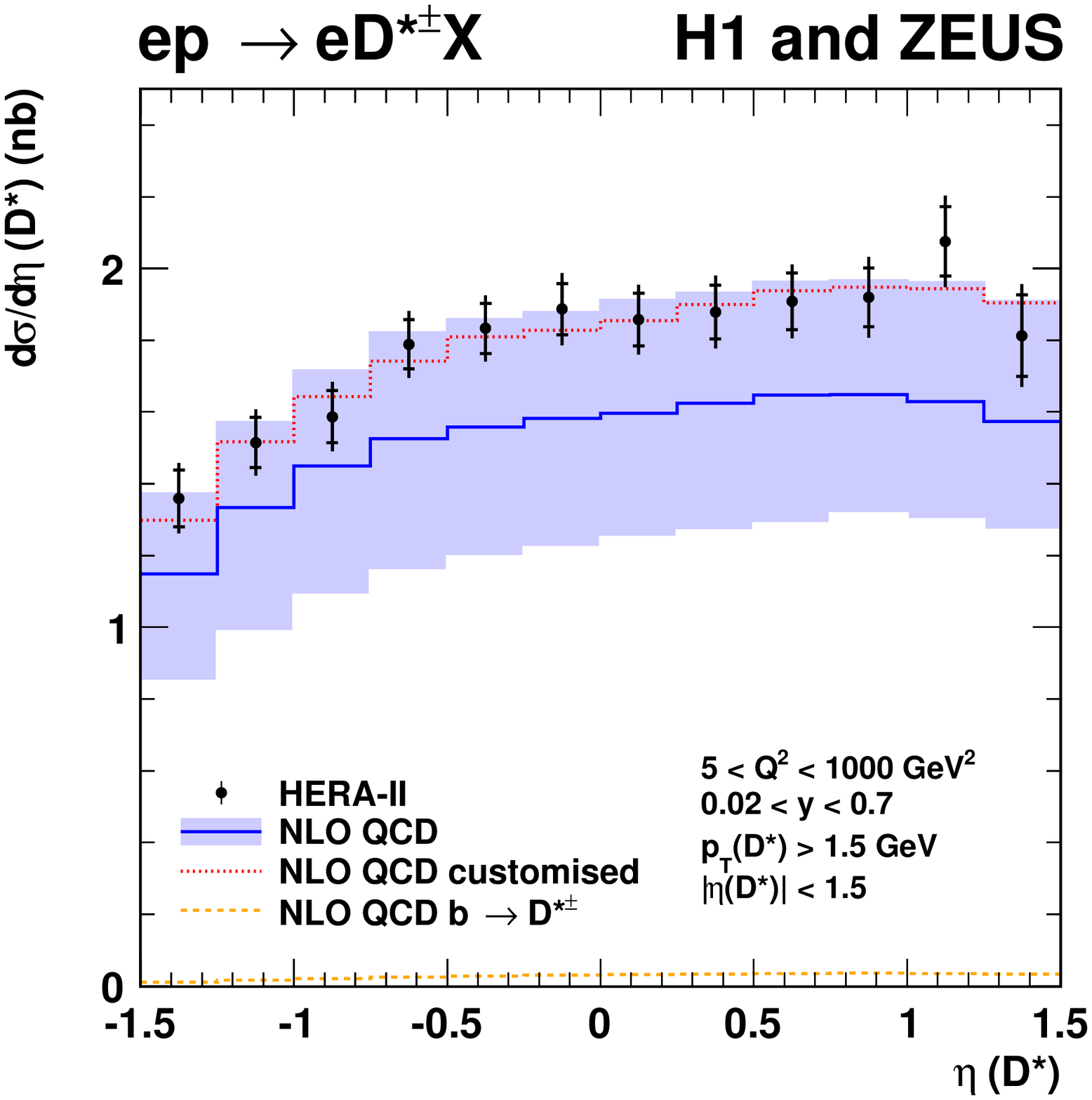,width=1.0\textwidth}
\caption{Differential $D^{*\pm}$-production cross section as a function of $\eta(D^{*})$.
The data points are the combined cross sections. The inner 
error bars indicate the uncorrelated part of the uncertainties.
The outer error bars represent the total uncertainties.
Also shown are the NLO predictions from HVQDIS (including the beauty 
contribution) and their uncertainty band.
A customised NLO calculation (dotted line, see text) is also shown.
}
\label{fig:etath} 
\end{figure}

\begin{figure}[h]
\center
\epsfig{file=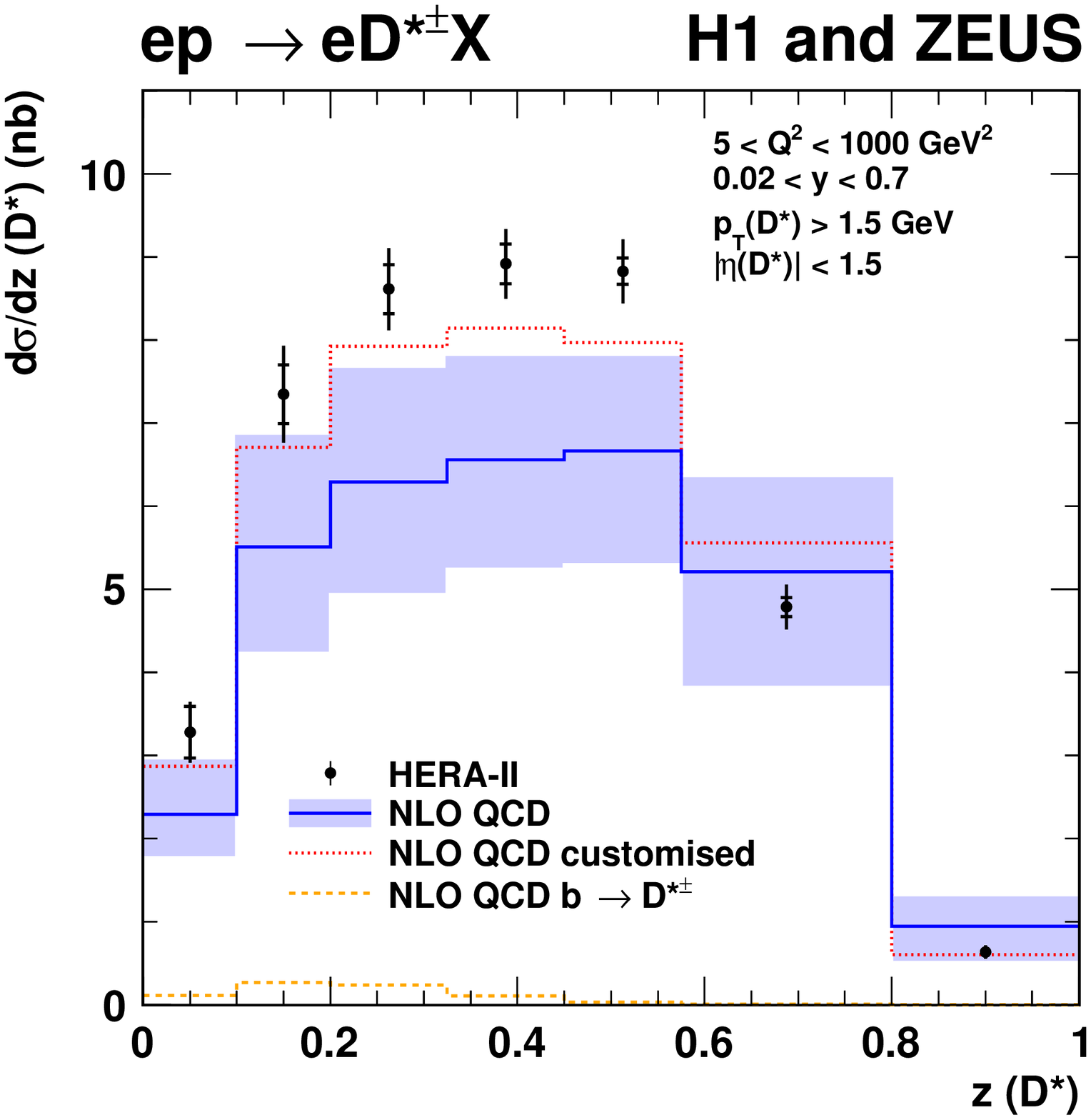,width=1.0\textwidth}
\caption{Differential $D^{*\pm}$-production cross section as a function of $z(D^{*})$.
The data points are the combined cross sections. The inner 
error bars indicate the uncorrelated part of the uncertainties.
The outer error bars represent the total uncertainties.
Also shown are the NLO predictions from HVQDIS (including the beauty 
contribution) and their uncertainty band.
A customised NLO calculation (dotted line, see text) is also shown.
}
\label{fig:zth} 
\end{figure}

\begin{figure}[h]
\center
\epsfig{file=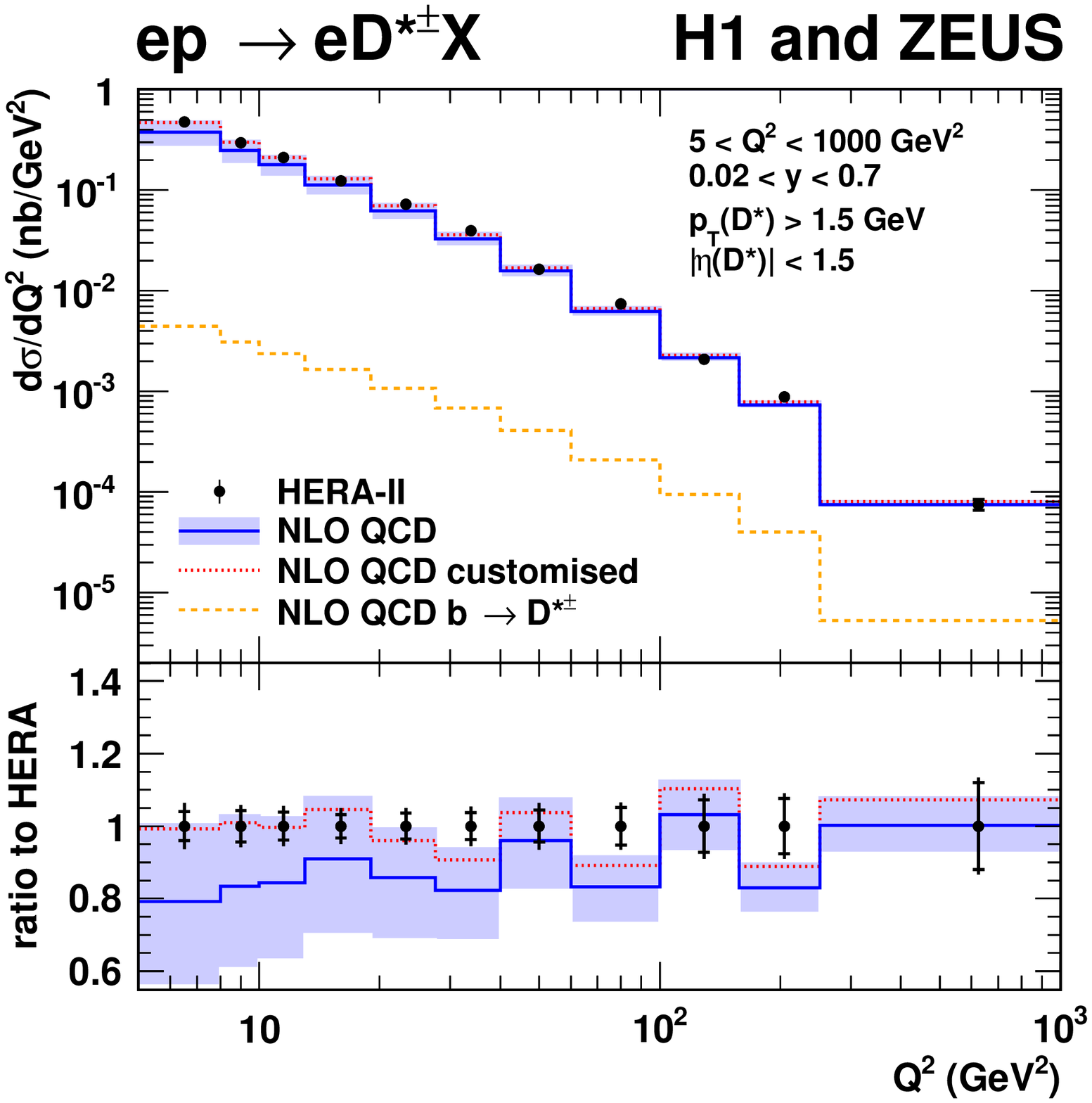,width=1.0\textwidth}
\caption{Differential $D^{*\pm}$-production cross section as a function of $Q^2$.
The data points are the combined cross sections. The inner 
error bars indicate the uncorrelated part of the uncertainties.
The outer error bars represent the total uncertainties.
Also shown are the NLO predictions from HVQDIS (including the beauty 
contribution) and their uncertainty band.
A customised NLO calculation (dotted line, see text) is also shown.
The bottom part shows the ratio of these cross sections with respect to the 
central value of the combined cross sections.}
\label{fig:q2th} 
\end{figure}

\begin{figure}[h]
\center
\epsfig{file=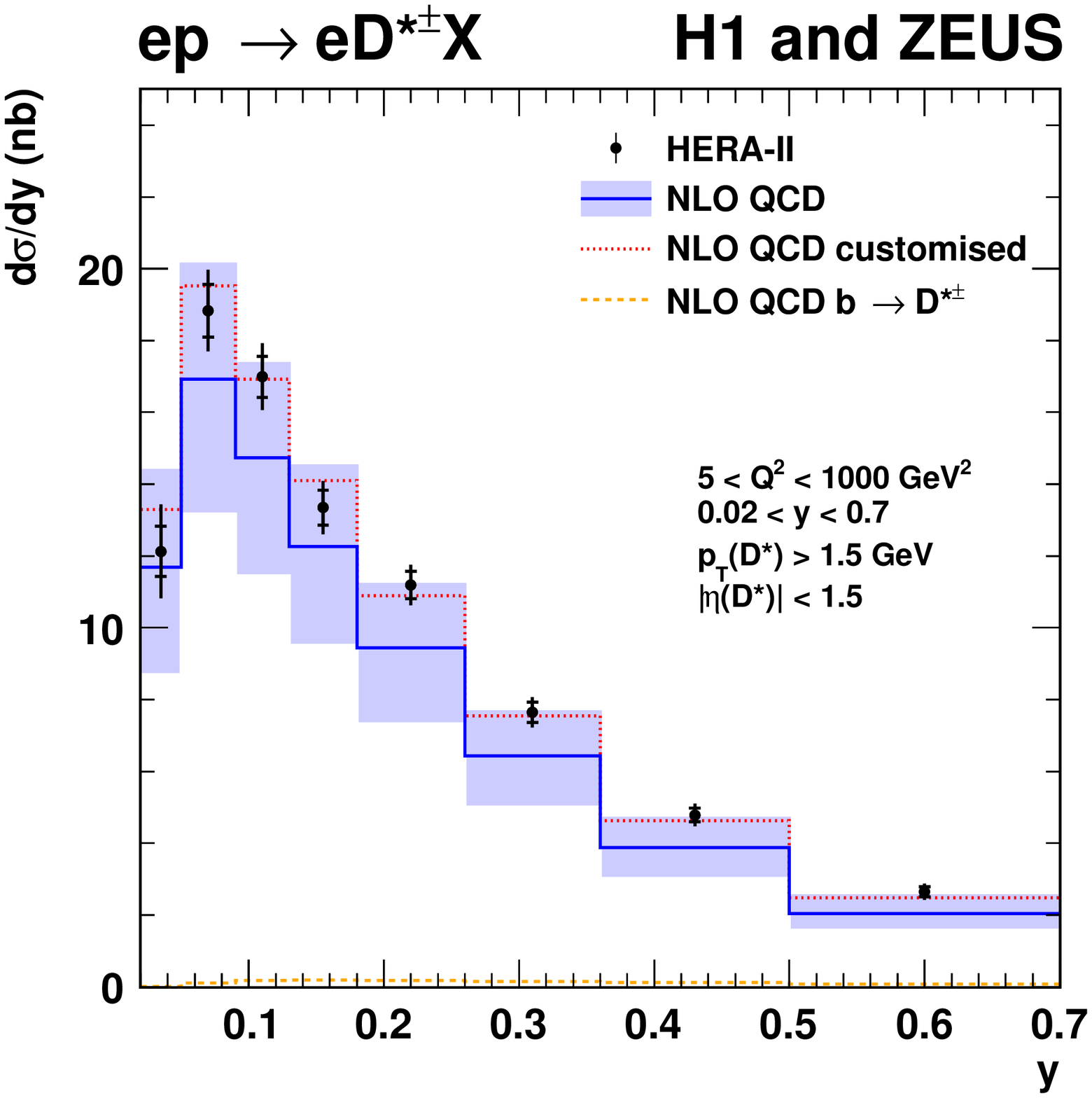,width=1.0\textwidth}
\caption{Differential $D^{*\pm}$-production cross section as a function of $y$.
The data points are the combined cross sections. The inner 
error bars indicate the uncorrelated part of the uncertainties.
The outer error bars represent the total uncertainties.
Also shown are the NLO predictions from HVQDIS (including the beauty 
contribution) and their uncertainty band.
A customised NLO calculation (dotted line, see text) is also shown.
}
\label{fig:yth} 
\end{figure}

The combined cross sections as a function of 
$p_T(D^*)$, $\eta(D^*)$, $z(D^*)$, $Q^2$ and $y$ 
are compared to NLO predictions%
\footnote{The NLO QCD prediction for the beauty contribution to $D^{*\pm}$ production, 
calculated as described in Section~\ref{sect:theory}, can be found on \\
\url{http://www.desy.de/h1zeus/dstar2015/}.}
in Figures~\ref{fig:ptth} -- \ref{fig:yth}.
In general, the predictions describe the data well. The data reach an 
overall precision of about 5\% over a large fraction of the measured 
phase space,
while the typical theoretical uncertainty ranges from 30\% at low $Q^2$ to 10\%
at high $Q^2$. The data points in the different distributions are 
statistically and systematically correlated. No attempt is made 
in this paper to quantify the correlations between bins taken from two 
different distributions. 
Thus quantitative comparisons of theory to data can only be made for individual distributions. 

\begin{figure}[h]
\center
\qquad\qquad\quad
\epsfig{file=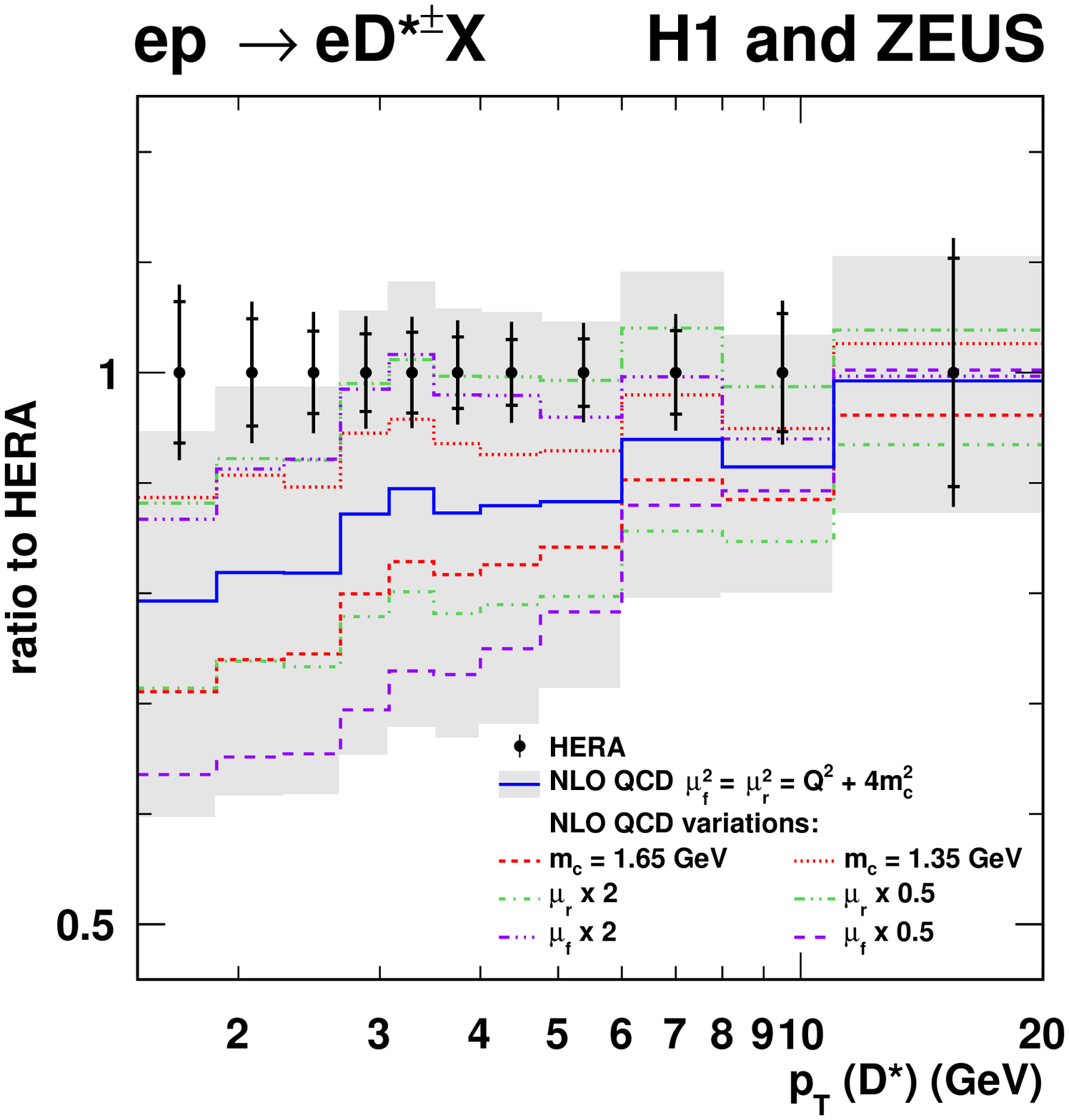,width=0.48\textwidth}
\vspace{.5cm}
\newline
\epsfig{file=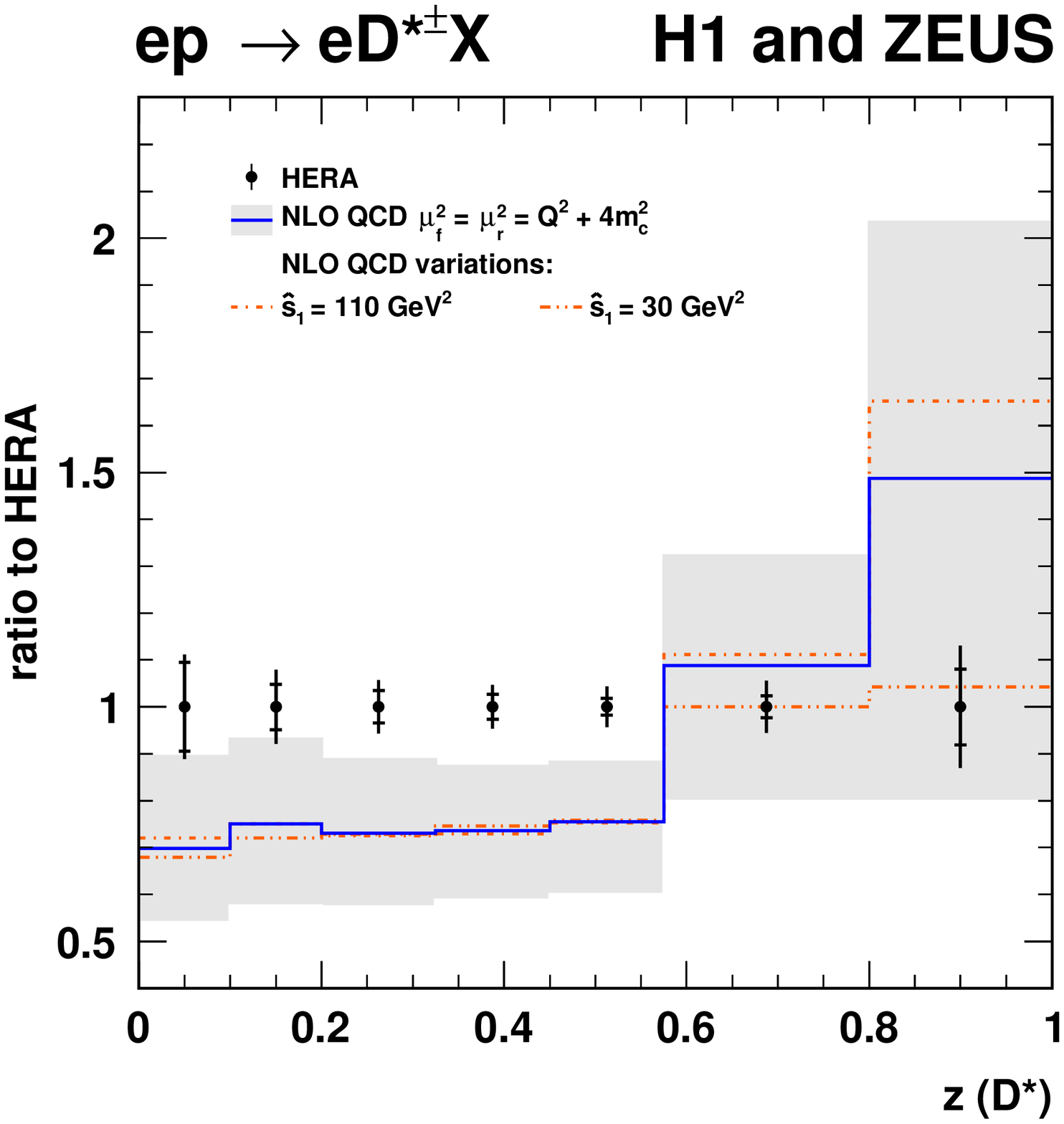,width=0.48\textwidth}
\epsfig{file=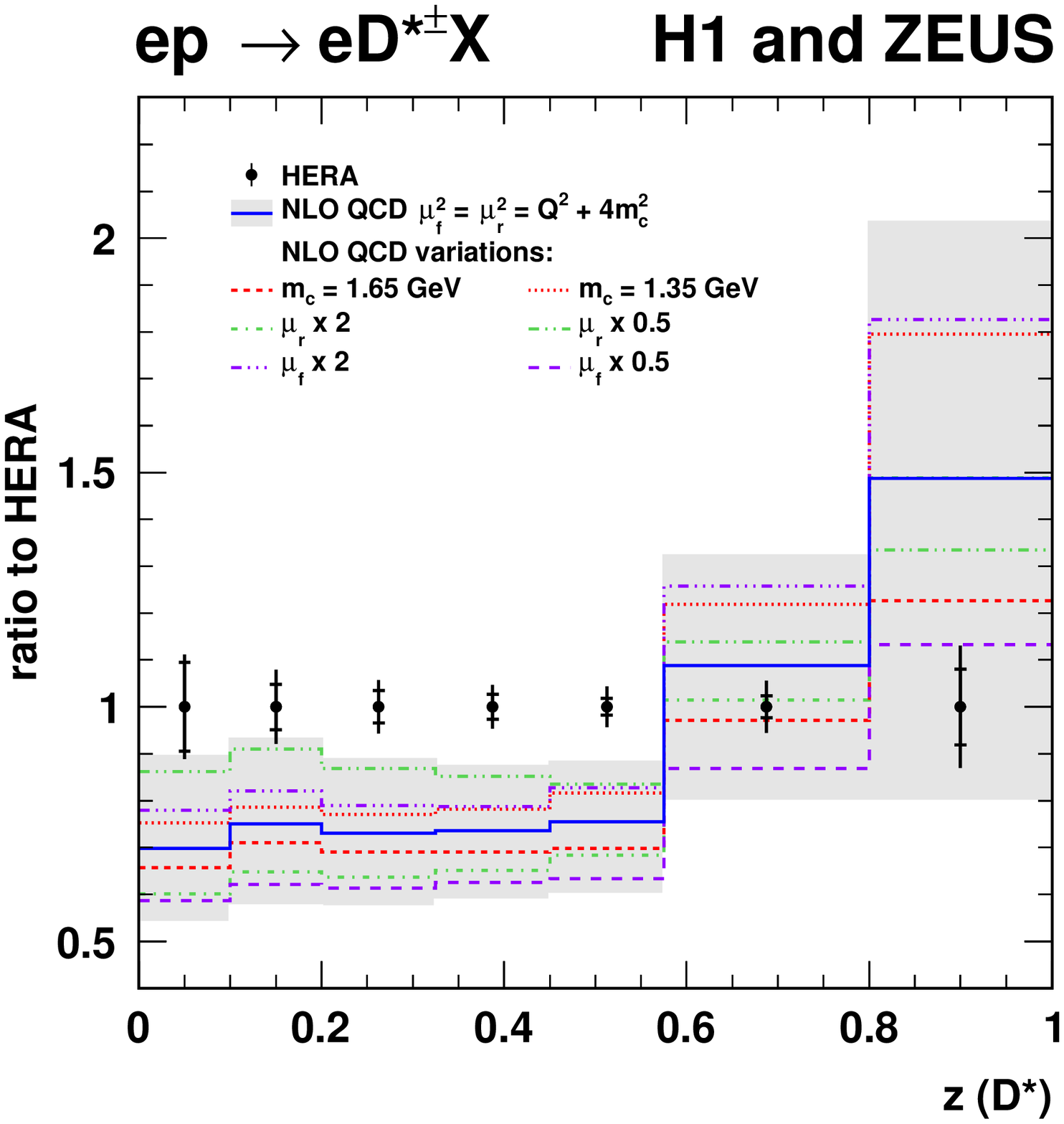,width=0.48\textwidth}
\caption{Differential $D^{*\pm}$-production cross section (ratio to data) as a function of $p_T(D^*)$ (top) 
and $z(D^*)$ (bottom) compared to NLO predictions with different variations:
charm-quark mass $m_{c}$, renormalisation scale $\mu_{r}$, 
factorisation scale $\mu_{f}$ and fragmentation bin boundary $\hat{s}_1$.}
\label{fig:thvars} 
\end{figure}

In order to study the impact of the current theoretical uncertainties in more 
detail, the effect of some variations on the predictions is shown separately 
in Figure~\ref{fig:thvars}, compared to the same data as in Figures~\ref{fig:ptth}, 
\ref{fig:zth} and \ref{fig:yth}. Only the variations with the largest 
impact on the respective distribution are shown in each case.

\begin{enumerate}
\item
The NLO prediction as a function of $p_T(D^*)$ (Figure~\ref{fig:thvars}, top) 
describes the data better by either
\begin{itemize}
\item setting the charm-quark pole mass to 1.35 GeV or
\item reducing the renormalisation scale by a factor 2 or
\item increasing the factorisation scale by a factor 2.
\end{itemize}
Simultaneous variation of both scales in the same direction would 
largely compensate and 
would therefore have a much smaller effect.  

  
\item
The prediction for the $z(D^*)$ distribution (Figure~\ref{fig:thvars}, bottom left) describes 
the shape of the data better if the fragmentation parameters 
are adjusted such that the boundary between the two lowest 
fragmentation 
regions \cite{HERAcharmcomb} is varied from the default of 70~GeV$^2$ to its 
lower boundary of 30~GeV$^2$. 

\item
The preference for a reduced renormalisation scale already observed for $p_T(D^*)$
is confirmed by the $z(D^*)$ distribution (Figure~\ref{fig:thvars}, bottom right). 
However, the shape of the $z(D^*)$ distribution rather prefers variations of 
the charm mass and the factorisation scale in the opposite direction to those 
found for the $p_T(D^*)$ distribution.
The distributions of the other kinematic variables do not provide additional 
information to these findings~\cite{ozthesis}.
\end{enumerate}

As stated before, within the large uncertainties indicated by the theory 
bands in Figures~\ref{fig:pt} -- \ref{fig:y}, all distributions are reasonably well 
described. However, the above study shows that the different contributions to 
these uncertainties do not only affect the normalisation, but also change 
the shape of different distributions in different ways. It is therefore 
not obvious that a variant of the prediction that gives a good description 
of the distribution in one variable will also give a good description 
of the distribution in another. 

Based on the study if items 1.-3. above, a `customised' calculation is 
performed to 
demonstrate the possibility of obtaining an improved description of 
the data in all variables at the same time, both in shape and normalisation,
within the theoretical uncertainties quoted in Section \ref{sect:theory}.
For this calculation, the following choices were made:
\begin{itemize}
\item From the three options discussed in item 1. above, the second is
    chosen, i.e. the renormalisation scale is reduced by a factor 2, with the 
factorisation scale unchanged.
\item The change of the fragmentation parameter $\hat s_1 = 30$~GeV$^2$, as discussed in item 2. above, is applied.
\item At this stage, the resulting distributions are still found to 
underestimate the data normalisation. 
As the renormalisation and factorisation scales are recommended to differ
by at most a factor of two \cite{scales}, the only
significant remaining handle arising from
items 1. and 3. is the charm-quark pole mass. This mass is set 
to 1.4 GeV, a value which is also compatible with the 
partially overlapping data used for a 
previous dedicated study \cite{HERAcharmcomb} of the charm-quark mass. 
\item All other parameters, which have a much smaller impact~\cite{ozthesis} 
than those discussed above, are left at their central settings as described in 
Section \ref{sect:theory}. 
\end{itemize} 

The result of this customised calculation is indicated as a dotted line in 
Figures~\ref{fig:ptth} -- \ref{fig:yth}. A reasonable agreement with 
data is achieved simultaneously in all variables. This a 
posteriori adjustment of theory parameters is not a prediction, 
but it can be taken as a hint in which direction theoretical and 
phenomenological developments could proceed. 
The strong improvement of the description of the data relative to the central 
prediction through the customisation of the renormalisation scale is in line 
with the expectation  
that higher-order calculations
will be helpful to obtain
a more stringent statement concerning the agreement of perturbative
QCD predictions with the data.
The improvement from the customisation of one of the fragmentation parameters
and the still not fully satisfactory description of the $z(D^*)$ distribution 
indicate that further dedicated experimental and theoretical studies of the 
fragmentation treatment
 might be helpful. 

In general, the precise single-differential distributions resulting from this 
combination, in particular those as a function of $p_T(D^*)$, $\eta(D^*)$ and $z(D^*)$, 
are sensitive to theoretical and phenomenological parameters in a way which 
complements the sensitivity of more inclusive variables like $Q^2$ and $y$.

\begin{figure}[h]
\center
\epsfig{file=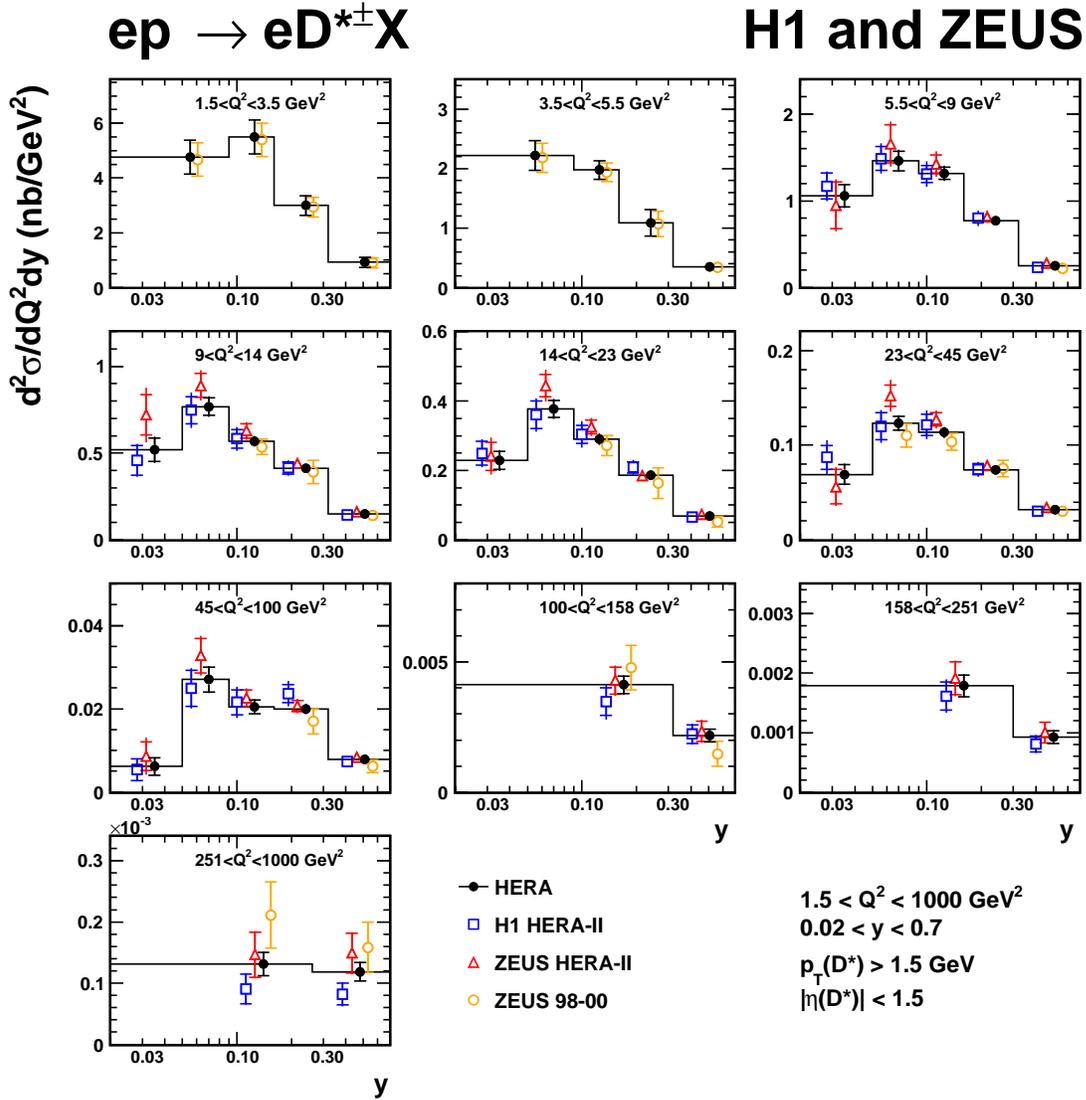,width=1.0\textwidth}
\caption{Double-differential $D^{*\pm}$-production 
cross sections as a function of $Q^2$ and $y$. 
The open triangles, squares and circles are the cross sections before combination, shown with a 
small horizontal offset for better visibility. 
The filled points are the combined cross sections. The inner 
error bars indicate the uncorrelated part of the uncertainties.
The outer error bars represent the total uncertainties.
}
\label{fig:q2y} 
\end{figure}

\begin{figure}[h]
\center
\epsfig{file=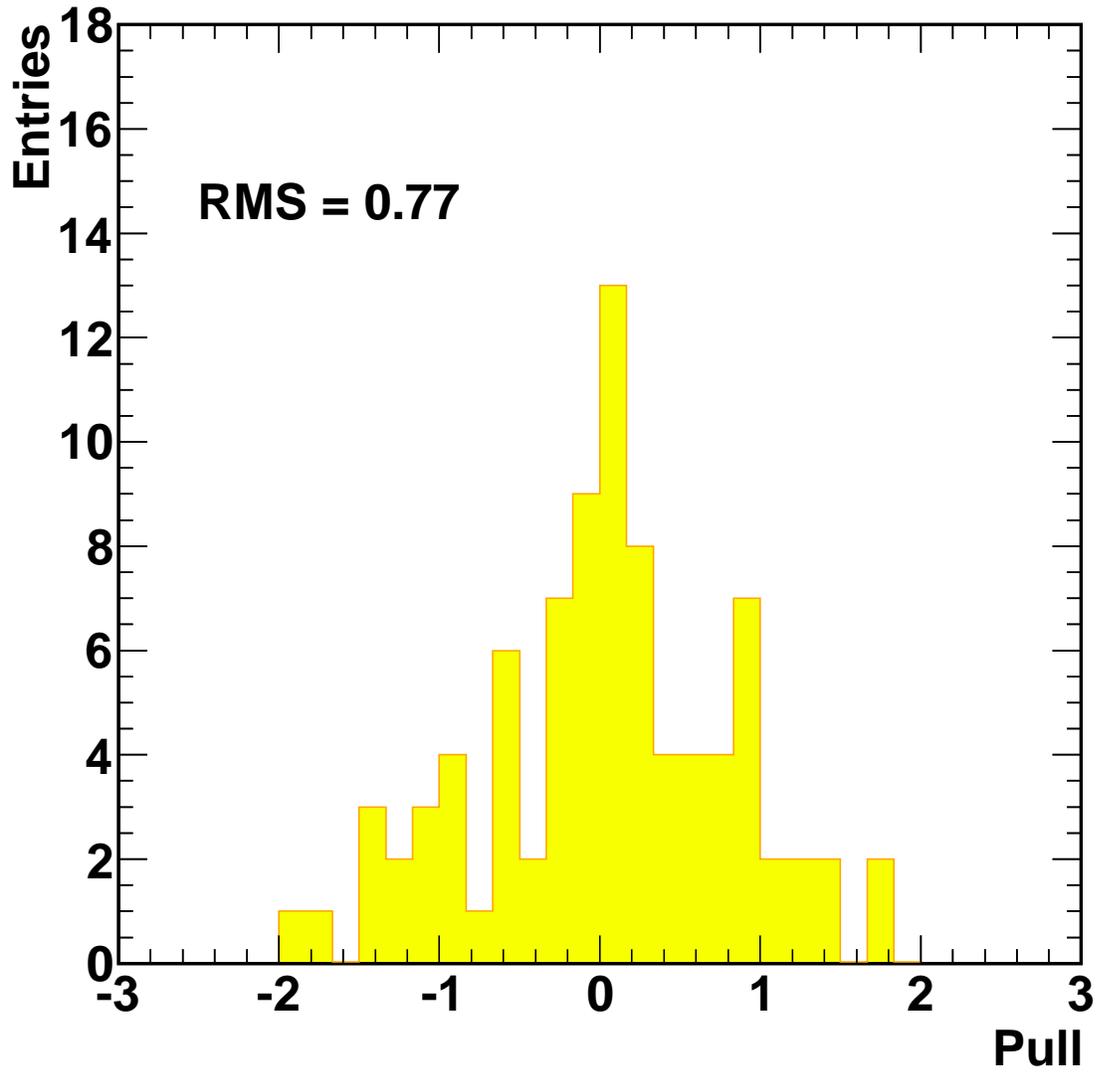,width=1.0\textwidth}
\caption{The pull distribution for the combination of the double-differential $D^{*\pm}$ cross sections.}
\label{fig:pull} 
\end{figure}

The combined double-differential cross sections with the uncorrelated, correlated
and total uncertainties\footnote{A detailed breakdown of correlated uncertainties can be 
found on 
\url{http://www.desy.de/h1zeus/dstar2015/}.}
as a function of $Q^2$ and $y$ are given
in Table~\ref{tab:result_q2y}. 
The total uncertainty is obtained by adding the uncorrelated and correlated uncertainties in quadrature.
The individual data sets as well as the results of the combination are shown
in Figure~\ref{fig:q2y}.
Including data set IV slightly reduces the overall 
cross-section normalisation with respect to the combination of 
sets I--III only.  
The pull distribution of the combination is shown in Figure~\ref{fig:pull}. 
The combination has a $\chi^2$ probability 
of 84\%, i.e. all data sets are consistent. 
The shifts and reductions of the correlated uncertainties are given in Table~\ref{tab:syst}. 

\begin{figure}[h]
\center
\epsfig{file=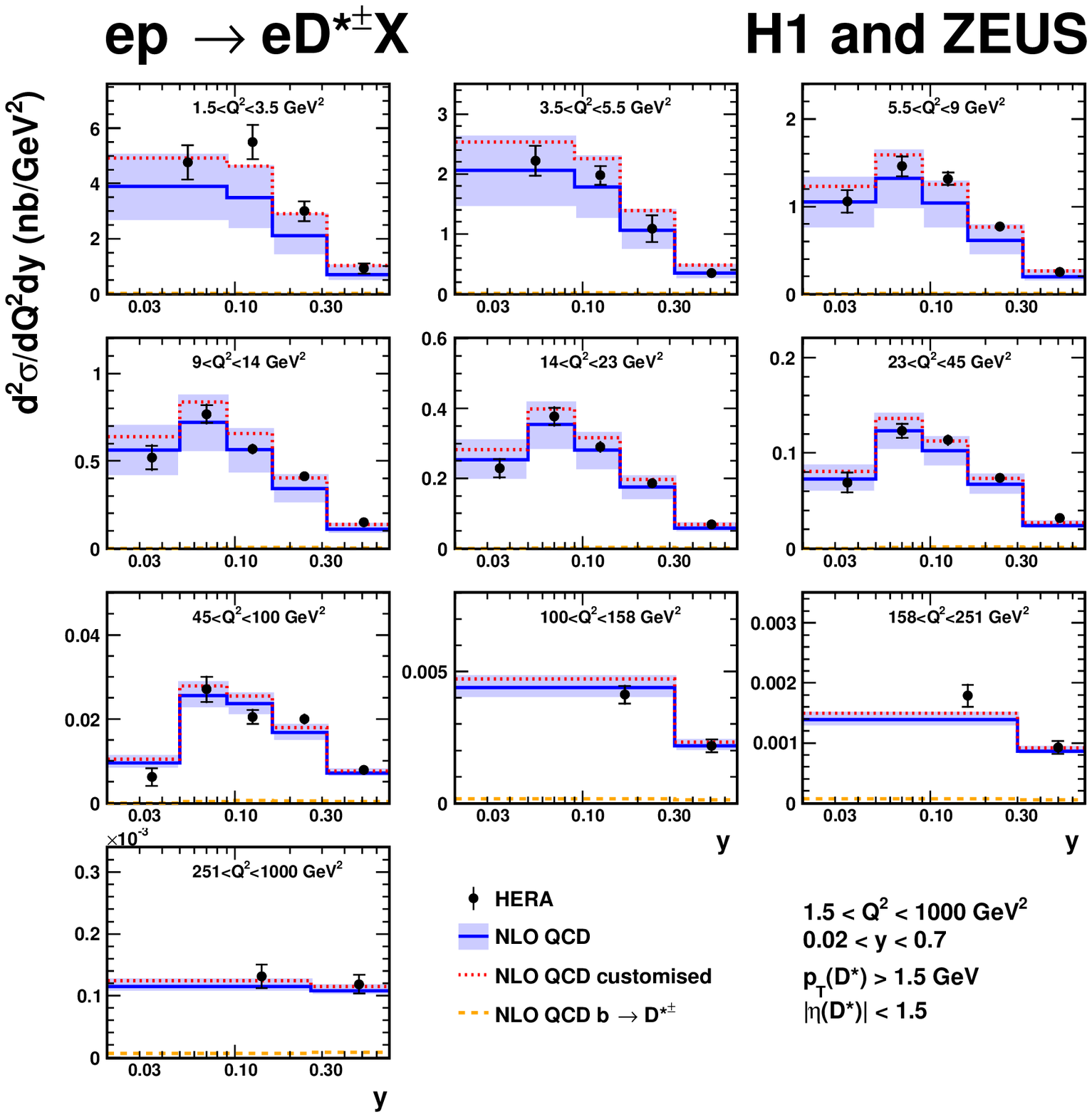,width=1.0\textwidth}
\caption{Double-differential $D^{*\pm}$-production cross section as a function 
of $Q^2$ and $y$.
The data points are the combined cross sections. The inner 
error bars indicate the uncorrelated part of the uncertainties.
The outer error bars represent the total uncertainties.
Also shown are the NLO predictions from HVQDIS (including the beauty 
contribution) and their uncertainty band.
A customised NLO calculation (dotted line, see text) is also shown.
}
\label{fig:q2yth} 
\end{figure}

These combined cross sections are compared to NLO predictions%
\footnote{The NLO QCD prediction for the beauty contribution to $D^{*\pm}$ production, 
calculated as described in Section~\ref{sect:theory}, can be found on \\
\url{http://www.desy.de/h1zeus/dstar2015/}.} in Figure~\ref{fig:q2yth}. 
The customised calculation is also shown. 
In general the predictions describe the data well. The data have a 
precision of about 5--10\% over a large fraction of the measured phase space,
while the estimated theoretical uncertainty ranges from 30\% at low $Q^2$ to 10\%
at high $Q^2$.
As well as the single-differential distributions, these double-differential distributions 
give extra input to test further theory improvements. 


\section{Conclusions}
\label{sect:conclude}
Measurements of $D^{*\pm}$-production cross sections in deep-inelastic $ep$ scattering by the 
H1 and ZEUS experiments are 
combined at the level of visible cross sections, 
accounting for their systematic correlations. 
The data sets were found to be consistent and the combined data have 
significantly reduced uncertainties. In contrast to the earlier charm 
combination at the level of reduced cross sections, 
the present combination does not have significant theory-related uncertainties 
and in addition distributions of kinematic variables of the $D^{*\pm}$ mesons 
are obtained. 
The combined data are compared to NLO QCD predictions.
The predictions describe the 
data well within their uncertainties. Higher order calculations
would be helpful to reduce the theory uncertainty to a level 
more comparable with the data precision. Further 
improvements in the treatment of heavy-quark fragmentation 
would also be desirable.

\section*{Acknowledgements}

We are grateful to the HERA machine group whose outstanding
efforts have made these experiments possible.
We appreciate the contributions to the construction, maintenance and operation of the H1 and ZEUS detectors of many people who are not listed as authors.
We thank our funding agencies for financial 
support, the DESY technical staff for continuous assistance and the 
DESY directorate for their support and for the hospitality they 
extended to the non-DESY members of the collaborations. 
We would like to give credit to all partners contributing to the EGI computing infrastructure for their support.

\noindent
\begin{flushleft}

\end{flushleft}

\newpage
\begin{table}[t]
\begin{center}
\tabcolsep2.1mm
\renewcommand*{\arraystretch}{1.2}
\begin{tabular}{|c|c|c|c|c|}
\hline
$p_T(D^{*})$ &$ \frac{{\rm d}\sigma}{{\rm d}p_T(D^{*})}$ &$\delta_\text{uncor}$  &$\delta_\text{cor}$&$\delta_\text{tot}$ \\
  (\gev)       & (nb/\gev)                         &$(\%)$            &$(\%)$ &$(\%)$ \\ 
\hline
$1.50$ : $1.88$ & $2.35$ & $6.4$ & $4.7$ & $8.0$ \\ \hline
$1.88$ : $2.28$ & $2.22$ & $4.9$ & $4.2$ & $6.4$ \\ \hline
$2.28$ : $2.68$ & $1.98$ & $3.7$ & $4.0$ & $5.5$ \\ \hline
$2.68$ : $3.08$ & $1.55$ & $3.5$ & $3.7$ & $5.1$ \\ \hline
$3.08$ : $3.50$ & $1.20$ & $3.7$ & $3.5$ & $5.1$ \\ \hline
$3.50$ : $4.00$ & $9.29 \times 10^{-1}$ & $3.2$ & $3.4$ & $4.7$ \\ \hline
$4.00$ : $4.75$ & $6.14 \times 10^{-1}$ & $3.0$ & $3.5$ & $4.6$ \\ \hline
$4.75$ : $6.00$ & $3.19 \times 10^{-1}$ & $3.1$ & $3.3$ & $4.5$ \\ \hline
$6.00$ : $8.00$ & $1.15 \times 10^{-1}$ & $3.8$ & $3.7$ & $5.3$ \\ \hline
$8.00$ : $11.00$ & $3.32 \times 10^{-2}$ & $5.4$ & $3.7$ & $6.5$ \\ \hline
$11.00$ : $20.00$ & $3.80 \times 10^{-3}$ & $10.4$ & $6.4$ & $12.2$ \\ \hline
\end{tabular}
\end{center}
\caption{The combined differential $D^{*\pm}$-production cross section in the phase space given in Table~\ref{Tab:vps} as a function of $p_T(D^{*})$, with its 
uncorrelated ($\delta_\text{uncor}$), correlated ($\delta_\text{cor}$) and total ($\delta_\text{tot}$) uncertainties.} 
\label{tab:result_pt}
\end{table}

\begin{table}[t]
\begin{center}
\tabcolsep2.1mm
\renewcommand*{\arraystretch}{1.2}
\begin{tabular}{|c|c|c|c|c|}
\hline
$\eta(D^{*})$ &$ \frac{{\rm d}\sigma}{{\rm d}\eta(D^{*})}$ &$\delta_\text{uncor}$  &$\delta_\text{cor}$&$\delta_\text{tot}$ \\
         & (nb)                         &$(\%)$            &$(\%)$ &$(\%)$ \\ 
\hline
$-1.50$ : $-1.25$ & $1.36$ & $5.8$ & $4.3$ & $7.2$ \\ \hline
$-1.25$ : $-1.00$ & $1.52$ & $4.6$ & $4.0$ & $6.1$ \\ \hline
$-1.00$ : $-0.75$ & $1.59$ & $4.6$ & $4.0$ & $6.1$ \\ \hline
$-0.75$ : $-0.50$ & $1.79$ & $3.8$ & $3.5$ & $5.2$ \\ \hline
$-0.50$ : $-0.25$ & $1.83$ & $3.8$ & $3.3$ & $5.1$ \\ \hline
$-0.25$ : $0.00$\ \ \ \ & $1.89$ & $3.8$ & $3.7$ & $5.3$ \\ \hline
$0.00$ : $0.25$ & $1.86$ & $4.0$ & $3.4$ & $5.2$ \\ \hline
$0.25$ : $0.50$ & $1.88$ & $4.0$ & $3.6$ & $5.4$ \\ \hline
$0.50$ : $0.75$ & $1.91$ & $4.1$ & $3.5$ & $5.4$ \\ \hline
$0.75$ : $1.00$ & $1.92$ & $4.3$ & $4.0$ & $5.9$ \\ \hline
$1.00$ : $1.25$ & $2.08$ & $4.7$ & $4.0$ & $6.1$ \\ \hline
$1.25$ : $1.50$ & $1.81$ & $6.3$ & $4.8$ & $7.9$ \\ \hline
\end{tabular}
\end{center}
\caption{The combined differential $D^{*\pm}$-production cross section in the phase space given in Table~\ref{Tab:vps} as a function of $\eta(D^{*})$, with its 
uncorrelated ($\delta_\text{uncor}$), correlated ($\delta_\text{cor}$) and total ($\delta_\text{tot}$) uncertainties.}
\label{tab:result_eta}
\end{table}

\begin{table}[t]
\begin{center}
\tabcolsep2.1mm
\renewcommand*{\arraystretch}{1.2}
\begin{tabular}{|c|c|c|c|c|}
\hline
$z(D^*)$ &$ \frac{{\rm d}\sigma}{{\rm d}z(D^{*})}$ &$\delta_\text{uncor}$  &$\delta_\text{cor}$&$\delta_\text{tot}$ \\
         & (nb)                         &$(\%)$            &$(\%)$ &$(\%)$ \\ 
\hline
$0.00$ : $0.10$ & $3.28$ & $9.5$ & $5.9$ & $11.2$ \\ \hline
$0.10$ : $0.20$ & $7.35$ & $4.8$ & $6.3$ & $7.9$ \\ \hline
$0.20$ : $0.32$ & $8.61$ & $3.5$ & $4.6$ & $5.7$ \\ \hline
$0.32$ : $0.45$ & $8.92$ & $2.7$ & $3.9$ & $4.7$ \\ \hline
$0.45$ : $0.57$ & $8.83$ & $1.8$ & $4.0$ & $4.3$ \\ \hline
$0.57$ : $0.80$ & $4.78$ & $2.4$ & $5.1$ & $5.6$ \\ \hline
$0.80$ : $1.00$ & $0.63$ & $8.1$ & $10.2$ & $13.0$ \\ \hline
\end{tabular}
\end{center}
\caption{The combined differential $D^{*\pm}$-production cross section in the phase space given in Table~\ref{Tab:vps} as a function of $z(D^{*})$, with its 
uncorrelated ($\delta_\text{uncor}$), correlated ($\delta_\text{cor}$) and total ($\delta_\text{tot}$) uncertainties.}
\label{tab:result_z}
\end{table}

\begin{table}[t]
\begin{center}
\tabcolsep2.1mm
\renewcommand*{\arraystretch}{1.2}
\begin{tabular}{|c|c|c|c|c|}
\hline
$Q^2$ &$ \frac{{\rm d}\sigma}{{\rm d}Q^2}$ &$\delta_\text{uncor}$  &$\delta_\text{cor}$&$\delta_\text{tot}$ \\
  (\gev$^{2}$)       & (nb/\gev$^{2}$)                         &$(\%)$            &$(\%)$ &$(\%)$ \\ 
\hline
$5$ : $8$ & $4.74 \times 10^{-1}$ & $4.0$ & $5.0$ & $6.4$ \\ \hline
$8$ : $10$ & $2.96 \times 10^{-1}$ & $4.3$ & $3.8$ & $5.8$ \\ \hline
$10$ : $13$ & $2.12 \times 10^{-1}$ & $3.8$ & $4.0$ & $5.6$ \\ \hline
$13$ : $19$ & $1.24 \times 10^{-1}$ & $3.2$ & $3.8$ & $5.0$ \\ \hline
$19$ : $28$ & $7.26 \times 10^{-2}$ & $3.5$ & $3.6$ & $5.0$ \\ \hline
$28$ : $40$ & $3.97 \times 10^{-2}$ & $3.7$ & $4.0$ & $5.5$ \\ \hline
$40$ : $60$ & $1.64 \times 10^{-2}$ & $4.4$ & $4.7$ & $6.4$ \\ \hline
$60$ : $100$ & $7.45 \times 10^{-3}$ & $5.2$ & $3.9$ & $6.5$ \\ \hline
$100$ : $158$ & $2.08 \times 10^{-3}$ & $7.2$ & $5.3$ & $9.0$ \\ \hline
$158$ : $251$ & $8.82 \times 10^{-4}$ & $7.6$ & $5.0$ & $9.1$ \\ \hline
$251$ : $1000$ & $7.50 \times 10^{-5}$ & $12.0$ & $6.7$ & $13.3$ \\ \hline
\end{tabular}
\end{center}
\caption{The combined differential $D^{*\pm}$-production cross section in the phase space given in Table~\ref{Tab:vps} as a function of $Q^2$, with its 
uncorrelated ($\delta_\text{uncor}$), correlated ($\delta_\text{cor}$) and total ($\delta_\text{tot}$) uncertainties.}
\label{tab:result_q2}
\end{table}

\begin{table}[t]
\begin{center}
\tabcolsep2.1mm
\renewcommand*{\arraystretch}{1.2}
\begin{tabular}{|c|c|c|c|c|}
\hline
$y$ &$ \frac{{\rm d}\sigma}{{\rm d}y}$ &$\delta_\text{uncor}$  &$\delta_\text{cor}$&$\delta_\text{tot}$ \\
         & (nb)                         &$(\%)$            &$(\%)$ &$(\%)$ \\ 
\hline
$0.02$ : $0.05$ & $12.1$ & $5.8$ & $9.1$ & $10.8$ \\ \hline
$0.05$ : $0.09$ & $18.8$ & $3.9$ & $4.6$ & $6.0$ \\ \hline
$0.09$ : $0.13$ & $17.0$ & $3.4$ & $4.3$ & $5.5$ \\ \hline
$0.13$ : $0.18$ & $13.4$ & $3.7$ & $4.2$ & $5.6$ \\ \hline
$0.18$ : $0.26$ & $11.2$ & $3.4$ & $3.7$ & $5.0$ \\ \hline
$0.26$ : $0.36$ & $7.65$ & $3.7$ & $4.2$ & $5.6$ \\ \hline
$0.36$ : $0.50$ & $4.78$ & $4.0$ & $5.3$ & $6.6$ \\ \hline
$0.50$ : $0.70$ & $2.65$ & $5.6$ & $6.4$ & $8.5$ \\ \hline
\end{tabular}
\end{center}
\caption{The combined differential $D^{*\pm}$-production cross section in the phase space given in Table~\ref{Tab:vps} as a function of $y$, with its 
uncorrelated ($\delta_\text{uncor}$), correlated ($\delta_\text{cor}$) and total ($\delta_\text{tot}$) uncertainties.}
\label{tab:result_y}
\end{table}

\begin{table}[tbp]
\small
\begin{center}
\tabcolsep0.55mm
\renewcommand*{\arraystretch}{1.2}
\begin{tabular}{|l|l|c|c|c|c|c|c|c|c|c|c|c|c|} \hline
\multirow{2}{*}{Data set} & \multirow{2}{*}{Name} & \multicolumn{2}{c}{$\frac{{\rm d}\sigma}{{\rm d}Q^2}$} & \multicolumn{2}{|c}{$\frac{{\rm d}\sigma}{{\rm d}y}$} & \multicolumn{2}{|c}{$\frac{{\rm d}\sigma}{{\rm d}p_T(D^{*})}$} & \multicolumn{2}{|c}{$\frac{{\rm d}\sigma}{{\rm d}\eta(D^{*})}$} & \multicolumn{2}{|c}{$\frac{{\rm d}\sigma}{{\rm d}z(D^{*})}$} & \multicolumn{2}{|c|}{$\frac{{\rm d}^2\sigma}{{\rm d}Q^2{\rm d}y}$} \\ \cline{3-14}
 &&sh&red&sh&red&sh&red&sh&red&sh&red&sh&red \\ \hline
I,II     &H1 CJC efficiency                  &   $0.8$&  $0.9$&   $0.3$&  $0.9$&   $0.5$&  $0.9$&   $0.5$&  $0.9$&   $0.4$&  $0.9$&   $0.6$&  $0.8$ \\ \hline
I,II     &H1 luminosity                      & $0.5$&  $0.9$&   $0.4$&  $0.9$&   $0.6$&  $0.9$&   $0.6$&  $0.9$&   $0.4$&  $0.9$&   $0.1$&  $0.9$ \\ \hline
I,II     & H1 MC PDF                         &  $0.1$&  $1.0$&   $0.1$&  $1.0$&   $0.2$&  $1.0$&   $0.2$&  $1.0$&   $0.1$&  $1.0$&   $0.0$&  $1.0$ \\ \hline
I,II     & H1 electron energy                &  $0.2$&  $1.0$&   $0.0$&  $1.0$&   $0.0$&  $1.0$&   $0.0$&  $1.0$&   $0.7$&  $0.9$&   $0.0$&  $0.8$ \\ \hline
I,II     & H1 electron polar angle           &  $0.2$&  $1.0$&   $0.1$&  $1.0$&   $0.1$&  $1.0$&   $0.2$&  $1.0$&   $0.2$&  $1.0$&   $0.3$&  $0.9$ \\ \hline
I,II     & H1 hadronic energy scale          &  $0.1$&  $1.0$&   $0.2$&  $0.9$&   $0.0$&  $1.0$&   $0.0$&  $1.0$&  $-1.0$&  $0.7$&   $0.0$&  $1.0$ \\ \hline
II       & H1 fragmentation threshold at high $Q^2$  &  $0.0$&  $1.0$&      &     &      &     &      &     &      &     &   $0.0$&  $1.0$ \\ \hline
I,II     & H1 alternative MC model           &  $0.4$&  $0.9$&   $0.4$&  $0.9$&   $0.1$&  $1.0$&   $0.0$&  $1.0$&  $-1.0$&  $0.8$&   $1.2$&  $0.7$ \\ \hline
I,II     & H1 alternative MC fragmentation           &  $0.0$&  $1.0$&   $0.0$&  $1.0$&   $0.0$&  $1.0$&  $-0.1$&  $1.0$&   $0.2$&  $1.0$&   $0.3$&  $0.9$ \\ \hline
I,II     & H1 fragmentation threshold        &  $0.0$&  $1.0$&  $-0.4$&  $0.9$&   $0.2$&  $1.0$&   $0.0$&  $1.0$&   $0.6$&  $0.9$&   $0.2$&  $0.8$ \\ \hline
I        & H1 high $Q^2$ uncertainty         &     &     &   $0.1$&  $1.0$&   $0.0$&  $0.9$&   $0.1$&  $1.0$&   $0.1$&  $1.0$&    &        \\ \hline
III       &  ZEUS hadronic energy scale     & $0.0$&  $1.0$&  $-0.1$&  $0.8$&   $0.0$&  $1.0$&   $0.0$&  $1.0$&  $-0.9$&  $0.9$&  $-0.5$&  $0.7$ \\ \hline
III       &   ZEUS electron energy scale    & $0.1$&  $0.9$&   $0.2$&  $0.9$&   $0.0$&  $1.0$&   $0.2$&  $1.0$&   $0.0$&  $1.0$&   $0.4$&  $0.7$ \\ \hline
III       &   ZEUS $p_T(\pi_s)$ correction   & $-0.1$&  $1.0$&  $-0.1$&  $1.0$&  $-0.1$&  $1.0$&  $-0.3$&  $1.0$&   $0.0$&  $1.0$&  $-0.7$&  $0.9$ \\ \hline
III       &   ZEUS $M(K\pi)$ window variation& $-0.3$&  $0.8$&  $-0.7$&  $0.8$&   $0.4$&  $0.6$&  $-0.3$&  $0.7$&   $0.5$&  $0.8$&  $-0.7$&  $0.9$ \\ \hline
III       &   ZEUS tracking efficiency      & $-0.2$&  $0.9$&  $-0.4$&  $0.9$&  $-0.4$&  $0.9$&  $-0.2$&  $0.9$&  $-0.2$&  $0.9$&  $-0.7$&  $1.0$ \\ \hline
III       &   ZEUS $b$ MC normalisation           & $0.0$&  $1.0$&   $0.0$&  $1.0$&   $0.0$&  $1.0$&   $0.0$&  $1.0$&   $0.1$&  $1.0$&   $0.0$&  $1.0$ \\ \hline
III       &   ZEUS PHP MC normalisation           & $0.0$&  $1.0$&  $-0.1$&  $1.0$&   $0.0$&  $1.0$&  $-0.1$&  $1.0$&   $0.1$&  $1.0$&  $-0.3$&  $1.0$ \\ \hline
III       &   ZEUS diffractive MC normalisation   & $0.0$&  $1.0$&   $0.1$&  $0.9$&   $0.2$&  $1.0$&   $0.0$&  $1.0$&   $0.0$&  $1.0$&   $0.7$&  $0.9$ \\ \hline
III       &    ZEUS MC reweighting ($p_T(D^{*})$ and $Q^2$)& $0.3$&  $0.9$&   $0.0$&  $1.0$&  $-0.1$&  $1.0$&   $0.0$&  $1.0$&   $0.0$&  $1.0$&   $0.6$&  $0.9$ \\ \hline
III       &   ZEUS MC reweighting ($\eta(D^{*})$)   & $0.0$&  $1.0$&   $0.0$&  $0.8$&  $-0.2$&  $1.0$&  $-0.3$&  $1.0$&  $-0.2$&  $1.0$&   $0.4$&  $0.8$ \\ \hline
III       &  ZEUS luminosity (HERA-II)      & $-0.2$&  $1.0$&  $-0.1$&  $1.0$&  $-0.2$&  $1.0$&  $-0.2$&  $1.0$&  $-0.1$&  $1.0$&  $-0.7$&  $0.9$ \\ \hline
IV       &	ZEUS luminosity (98-00)          &   &     &      &     &      &     &      &     &      &     &  $0.8$ &  $0.9$ \\ \hline
I-IV       &        Theory $m_c$ variation            & &   &     &    &     &    &     &    &     &    &    $0.0$&  $1.0$ \\ \hline
I-IV       &        Theory $\mu_r$, $\mu_f$ variation & &   &     &    &     &    &     &    &     &    &    $0.0$&  $1.0$ \\ \hline
I-IV       &        Theory $\alpha_s$ variation       & &   &     &    &     &    &     &    &     &    &    $0.0$&  $1.0$ \\ \hline
I-IV       &        Theory longitudunal frag. variation         & &    &     &    &     &    &     &    &     &    &    $0.1$&  $1.0$ \\ \hline
I-IV       &        Theory transverse frag. variation           & &    &     &    &     &    &     &    &     &    &    $0.0$&  $1.0$ \\ \hline
\end{tabular}
\end{center}
\caption[Sources of  correlated uncertainties in combination of $D^{*\pm}$ visible cross sections]
{Sources of point-to-point correlated uncertainties. 
  For each source the affected data sets are given, together with the shift (sh) and reduction factor (red) in the combination obtained after the first iteration. 
  For sources which do not affect the combination of a given differential cross section, no shifts and reductions are quoted.
  }
\label{tab:syst}
\end{table}

\begin{table}[t]
\begin{center}
\tabcolsep2.1mm
\renewcommand*{\arraystretch}{1.03}
\begin{tabular}{|c|c|c|c|c|c|}
\hline
$Q^2$ & $y$ & $ \frac{{\rm d}^2\sigma}{{\rm d}Q^2{\rm d}y}$ &$\delta_\text{uncor}$  &$\delta_\text{cor}$&$\delta_\text{tot}$ \\
  (\gev$^{2}$)  &     & (nb/\gev$^{2}$)                         &$(\%)$            &$(\%)$ &$(\%)$ \\ 
\hline
$1.5$ : $3.5$ & $0.02$ : $0.09$ & $4.76$ & $12.9$ & $2.5$ & $13.2$ \\
        & $0.09$ : $0.16$ & $5.50$ & $11.3$ & $2.6$ & $11.5$ \\
        & $0.16$ : $0.32$ & $3.00$ & $12.0$ & $2.6$ & $12.3$ \\
        & $0.32$ : $0.70$ & $9.21 \times 10^{-1}$ & $20.5$ & $2.5$ & $20.7$ \\ \hline
$3.5$ : $5.5$ & $0.02$ : $0.09$ & $2.22$ & $11.3$ & $2.8$ & $11.6$ \\
        & $0.09$ : $0.16$ & $1.98$ & $7.9$ & $2.7$ & $8.3$ \\
        & $0.16$ : $0.32$ & $1.09$ & $20.2$ & $2.7$ & $20.4$ \\
        & $0.32$ : $0.70$ & $3.47 \times 10^{-1}$ & $14.6$ & $2.6$ & $14.8$ \\ \hline
$5.5$ : $9$ & $0.02$ : $0.05$ & $1.06$ & $12.3$ & $4.4$ & $13.1$ \\
        & $0.05$ : $0.09$ & $1.46$ & $7.8$ & $4.1$ & $8.8$ \\
        & $0.09$ : $0.16$ & $1.32$ & $5.4$ & $4.3$ & $6.9$ \\
        & $0.16$ : $0.32$ & $7.73 \times 10^{-1}$ & $4.9$ & $3.9$ & $6.3$ \\
        & $0.32$ : $0.70$ & $2.51 \times 10^{-1}$ & $5.6$ & $4.2$ & $7.0$ \\ \hline
$9$ : $14$ & $0.02$ : $0.05$ & $5.20 \times 10^{-1}$ & $13.0$ & $6.6$ & $14.6$ \\
        & $0.05$ : $0.09$ & $7.68 \times 10^{-1}$ & $6.6$ & $3.9$ & $7.7$ \\
        & $0.09$ : $0.16$ & $5.69 \times 10^{-1}$ & $4.6$ & $2.8$ & $5.4$ \\
        & $0.16$ : $0.32$ & $4.12 \times 10^{-1}$ & $4.6$ & $3.1$ & $5.6$ \\
        & $0.32$ : $0.70$ & $1.51 \times 10^{-1}$ & $5.6$ & $4.0$ & $6.9$ \\ \hline
$14$ : $23$ & $0.02$ : $0.05$ & $2.29 \times 10^{-1}$ & $11.4$ & $6.3$ & $13.0$ \\
        & $0.05$ : $0.09$ & $3.78 \times 10^{-1}$ & $6.5$ & $4.1$ & $7.7$ \\
        & $0.09$ : $0.16$ & $2.90 \times 10^{-1}$ & $4.8$ & $3.3$ & $5.8$ \\
        & $0.16$ : $0.32$ & $1.86 \times 10^{-1}$ & $5.0$ & $3.4$ & $6.0$ \\
        & $0.32$ : $0.70$ & $6.92 \times 10^{-2}$ & $6.2$ & $4.4$ & $7.7$ \\ \hline
$23$ : $45$ & $0.02$ : $0.05$ & $6.91 \times 10^{-2}$ & $14.8$ & $8.2$ & $16.7$ \\
        & $0.05$ : $0.09$ & $1.23 \times 10^{-1}$ & $5.9$ & $3.6$ & $6.9$ \\
        & $0.09$ : $0.16$ & $1.14 \times 10^{-1}$ & $4.4$ & $3.0$ & $5.3$ \\
        & $0.16$ : $0.32$ & $7.42 \times 10^{-2}$ & $4.3$ & $3.0$ & $5.2$ \\
        & $0.32$ : $0.70$ & $3.21 \times 10^{-2}$ & $5.2$ & $3.7$ & $6.4$ \\ \hline
$45$ : $100$ & $0.02$ : $0.05$ & $6.16 \times 10^{-3}$ & $33.5$ & $11.1$ & $35.3$ \\
        & $0.05$ : $0.09$ & $2.70 \times 10^{-2}$ & $11.0$ & $4.4$ & $11.8$ \\
        & $0.09$ : $0.16$ & $2.05 \times 10^{-2}$ & $8.0$ & $3.7$ & $8.8$ \\
        & $0.16$ : $0.32$ & $1.99 \times 10^{-2}$ & $5.4$ & $3.2$ & $6.3$ \\
        & $0.32$ : $0.70$ & $7.84 \times 10^{-3}$ & $6.9$ & $4.0$ & $7.9$ \\ \hline
$100$ : $158$ & $0.02$ : $0.32$ & $4.12 \times 10^{-3}$ & $8.2$ & $4.1$ & $9.2$ \\
        & $0.32$ : $0.70$ & $2.18 \times 10^{-3}$ & $11.1$ & $4.1$ & $11.9$ \\ \hline
$158$ : $251$ & $0.02$ : $0.30$ & $1.79 \times 10^{-3}$ & $10.2$ & $4.4$ & $11.1$ \\
        & $0.30$ : $0.70$ & $9.28 \times 10^{-4}$ & $11.6$ & $4.6$ & $12.5$ \\ \hline
$251$ : $1000$ & $0.02$ : $0.26$ & $1.31 \times 10^{-4}$ & $14.5$ & $4.7$ & $15.3$ \\
        & $0.26$ : $0.70$ & $1.18 \times 10^{-4}$ & $12.7$ & $5.0$ & $13.6$ \\ \hline
\end{tabular}
\end{center}
\caption{The combined double-differential $D^{*\pm}$-production cross section in the phase space given in Table~\ref{Tab:vps} as a function of $Q^2$ and $y$, with its 
uncorrelated ($\delta_\text{uncor}$), correlated ($\delta_\text{cor}$) and total ($\delta_\text{tot}$) uncertainties.}
\label{tab:result_q2y}
\end{table}


\newpage

\newpage

\newpage

\newpage

\newpage

\newpage

\newpage

\newpage

\newpage

\newpage

\newpage

\newpage

\newpage

\newpage


\unitlength1cm
\end{document}